\documentclass[%
 reprint,
 amsmath,amssymb,
 aps,
onecolumn,
]{revtex4-2}

\usepackage{graphicx}
 \usepackage{dcolumn}
\usepackage{bm}
\usepackage{xcolor}
\usepackage{color} 
\usepackage{verbatim}
\usepackage{amsthm}
\usepackage{subfig}

\newtheorem{remark}{Remark}  

\graphicspath{{FIG/}}

\begin{document}

\preprint{APS/123-QED}

\title{Upscaling the Navier-Stokes-Cahn-Hilliard model  for incompressible multiphase flow in inhomogeneous porous media}
 \author{Chunhua Zhang}
\affiliation{
School of Energy and Power Engineering, North University of China, Taiyuan, Shanxi, 030051, China}
\affiliation{
State Key Laboratory of Coal and CBM Co-Mining, North University of China, Taiyuan, Shanxi 030051, China}
\author{Peiyao Liu}
\affiliation{
College of Mathematics, Taiyuan University of Technology,  Taiyuan, Shanxi, 030024, China}
\author{Cheng Peng}
\email{pengcheng@sdu.edu.cn}
\affiliation{
Key Laboratory of high efficiency and clean mechanical manufacture ministry of education school of mechanical engineering,
Shandong University, Jinan, Shandong, 250061, China}
\author{Lian-Ping Wang}
\affiliation{
Guangdong Provincial Key Laboratory of Turbulence Research and Applications, Department of Mechanics and Aerospace Engineering, Southern University of Science
 and Technology, Shenzhen 518055, Guangdong, China}
\author{Zhaoli Guo}
\email{zlguo@hust.edu.cn}
\affiliation{
Institute of Interdisciplinary Research for Mathematics and Applied Science,  Huazhong University of Science and Technology, Wuhan 430074, China}

\date{\today}

\begin{abstract}
This work presents a macroscopic model for the flow of two immiscible and incompressible fluids within  inhomogeneous porous media. At the pore scale, the flow is governed by the full Navier-Stokes equations while the phase interface evolution is described by the Cahn-Hilliard equation. Applying the volume averaging method, we rigorously derive upscaled equations that characterize the Darcy-scale behavior of the two-phase system. The derivation yields unclosed terms originating from spatial derivations, which are subsequently closed by modeling them as functions of averaged quantities and specific transport coefficients.
These coefficients are evaluated by solving localized closure problems defined on representative elementary volumes (REVs). A key contribution of this study is the formal incorporation of wetting behavior into the averaged chemical potential.  We further discuss the theoretical distinctions between the proposed framework and standard empirical two-phase Darcy models. Finally, numerical simulations of the upscaled equations are performed, demonstrating the model’s capability to capture essential two-phase flow characteristics in porous media.
 
\end{abstract}

\maketitle


\section{Introduction}
Two-phase flows through porous media occur in a wide variety of engineering  applications such as groundwater remediation, enhanced oil recovery,  $\text{CO}_2$ sequestration,  and many others.
In such situations, the flow and transport processes take place at the scale of pores.
Since the geometry of the porous media is highly complex, the modeling of multiphase flow in the porous media is rather challenging.
The direct numerical simulation (DNS)  that solves the complete governing equations at the pore scale plays an important role to understand the underlying transport mechanism.
Using the DNS for flow in various porous media,  many important pore-scale flow phenomena  have been investigated, including
pore-scale displacement~\cite{lenormand1988numerical,singh2017dynamics,singh2019capillary},  spontaneous imbibition~\cite{liu2022systematic,lin2021spontaneous,gu2021preferential}, wetting hysteresis~\cite{joekar2013trapping,qin2021lattice,shams2021direct}, pore structure properties~\cite{tang2021review,zakirov2020prediction}, relative permeability~\cite{keehm2004permeability,shi2018relative,zhao2018effect}  and so on.
Several phase diagrams of flow regimes have been identified based on the macroscopic displacement patterns~\cite{lenormand1988numerical,zhao2016wettability,primkulov2021wettability}.
Although these findings of the flow behaviors at the pore scale are important, it is still difficult to  deal with such flow in the engineering due to its wide ranges of time scales and space scales.
The averaged behaviours of the two-phase system at a much larger scale (also known as Darcy-scale or REV) is required to optimize  technological  processes and improve industrial production.



At the REV scale, the fundamental law governing the flow of fluids through porous media is  Darcy's law.
This law represents an empirical formulation originally deduced from experimental studies of one-dimensional single-phase flow.
Despite this,  the Darcy's law has been widely and successfully applied over the decades to model and simulate various flow phenomena.
However, it was soon recognized that \textcolor{red}{Darcy's law fails to accurately describe the flow behavior  under high-velocity or turbulent conditions, where
the relationship between the velocity flux  and the pressure gradient becomes non-linear}.

To enhance the accuracy of the Darcy's law, many modified approaches have been proposed. For example,  Brinkman~\cite{brinkman1949calculation} added  second-order velocity derivatives  into the Darcy equation, leading to the Brinkman equation. Similarly, Dupuit and  Forchheimer ~\cite{ph1901wasserbewegung}  incorporated a quadratic term of the velocity to account for microscopic inertial effect, resulting in the Brinkman-Forchheimer equation.
Zeng et al.~\cite{zeng2006criterion} further advanced this equation by proposing a revised Forchheimer number as a criterion for identifying non-Darcy flow in porous media.

when considering flow of  multiple fluids in a porous medium,  the aforementioned models have been extended to describe multiphase flow in porous media~\cite{whitaker1986flow,dullien1988two,niessner2011comparison}.
These extended formulations closely resemble the  single-phase Darcy formulation, but
 several key parameters related to the interfacial properties  have been introduced,  including phase saturation,   the relative permeability and the capillary pressure  between the two phases. These additions highlight the complexity involved in parameterizing Darcy's law for the multiphase flow situation.
To fully close the two-phase Darcy's law, many empirical functions for relative permeability and capillary pressure have been developed by
performing a reasonably good fit to experimental data.
In most formulations, the relative permeability and  the capillary pressure are assumed to depend solely on phase saturation.
However, it has been observed that the dependence of both the relative permeability and the capillary pressure  on saturation  exhibits significant hysteretic behavior.
Specifically, their values are not uniquely determined by saturation alone; rather, they also depend  on the displacement history, such as whether the process involves  drainage or imbibition. This suggests that the standard parameterization of  two-phase Darcy flow is insufficient to capture the underlying physics.

To reconcile the discrepancies between  experimental measurements and predictions of empirical formulations,numerious improved  empirical and semi-empirical functions have been developed by incorporating  key factors, such as rock wettability~\cite{guo2022role},
fractal dimension of pore size~\cite{xu2013prediction}, temperature~\cite{torabi2016predicting}, interfacial tension~\cite{suwandi2022relative}, fluid viscosity~\cite{esmaeili2019review}, phase distribution~\cite{khorsandi2017equation} and others.
\textcolor{red}{Recently, Purswani et al~\cite{purswani2020development} proposed an equation of state for relative permeability based on  predetermined state variables, whose calibrating parameters could be determined via linear  regression on the experimental or pore-scale data.
Because the Darcy's law for both single fluid and two-phase fluids was originally proposed based on experiment rather than derived rigorously from the physical governing equations, the selection of state parameters for different flow regimes often relies on  personal experience and knowledge.
This indicates that a rational and systematic understanding of how microscopic details affect macroscopic flow behavior is essential for  building  appropriate equations at the Darcy scale or for  modifying relative permeabiity.
Consequently, there is a growing recognition that a more fundamental approach that can integrate pore-level dynamics into Darcy-scale  descriptions is needed to overcome the limitations of empirical corrections. 
In view of this, the upscaling of multiphase flow at the pore scale has been a focus of much attention.}

To date, the upscaling of transport and flow processes in porous media  from the pore-scale to the macroscopic scale has primarily relied on the volume averaging method~\cite{whitaker1998method,gray1977theorems} and  multiscale asymptotics (also referred to as homogenization theory) ~\cite{pavliotis2008multiscale,mei2010homogenization,hassani1998review}.  Both approaches originate from a continuum mechanics-based model.

In the multiscale asymptotics method, physical quantities are formulated as functions of two distinct spatial scales, a amacroscopic scale, characterizing the gradual variations across the domain, and a microscopic scale, capturing the rapid fluctuations within the pore structure. \textcolor{red}{By employing a multiscale expansion and invoking periodicity assumptions on the unit cell, effective upscaled equations and homogenized parameters are rigorously derived. }
For example,
Sharmin et al ~\cite{sharmin2022upscaling} derived a two-scale model describing the averaged behaviour of the flow of two immiscible and incompressible fluid phases in a porous medium.

Conversely, the volume averaging method operates by integrating governing equations over a REV. The REV must be sufficiently large to smooth out pore-level irregularities, yet adequately small to resolve macroscopic gradients and capture non-equilibrium effects. \textcolor{red}{A fundamental distinction lies in the treatment of spatial deviations. Within the volume averaging framework, physical quantities are decomposed into an averaged component and a spatial deviation term. 
Unlike the multiscale asymptotics expansions, where the Fredholm alternative inherently ensures existence and uniqueness, the volume-averaged equations do not form a closed system due to the absence of direct governing equations for the deviation terms. Consequently, closure must be achieved through constitutive assumptions or the solution of associated closure problems within the REV.
Despite its inherent complexity, the volume averaging method is often favored for its  physical transparency, namely, the averaging procedure yields variables with clear physical interpretations.  As noted by Davit et al.~\cite{davit2013homogenization}, while multiscale asymptotics offers superior mathematical rigor for periodic structures, volume averaging provides a more flexible and intuitive framework for practical applications in complex, non-periodic porous media. }

The utility of this approach was famously demonstrated by Whitaker, who derived   Darcy’s Law from the Stokes~\cite{whitaker1986flow} and Navier-Stokes equations without initial constitutive assumptions~\cite{whitaker1996forchheimer}. \textcolor{red}{Building on this foundation, Chen et al.~\cite{chen2019homogenization} recently utilized local volume averaging to upscale the Stokes-Cahn-Hilliard system for two-phase incompressible flow.
Nevertheless, the derived two-phase Darcy formulation constitutes a two-fluid approach, characterized by distinct velocities and separate pressure fields. In particular, 
the averaged chemical potential was not explicitly derived but was directly substituted by an empirical capillary pressure function, thereby masking the influence of pore-scale information.}


\textcolor{blue}{This study aims to derive a macroscopic single-fluid model for two-phase flow in  porous media by upscaling  pore-scale governing equations via the volume averaging method.}
Among various frameworks for modeling multiphase flow,  the Navier-Stokes-Cahn-Hilliard (NSCH) system is  employed  due to its rigorous thermodynamic foundation. 
The model is derived from  a free energy functional, which can provide a unified and thermodynamically consistent description of phase behavior. \textcolor{blue}{
Crucially,  we focus on the rigorous derivation of capillary effects and wetting behavior  from the pore-scale surface tension force and  boundary conditions.
By applying the spatial  averaging theorems, the wetting boundary condition is directly incorporated into the averaged chemical potential, which subsequently governs the  capillary effects.
Through judicious simplifications, 
we   obtain a set of partial differential equations governing the volume-averaged quantities.}
To evaluate the capability of the proposed model,  we utilize the phase-field-based lattice Boltzmann method  to solve the derived averaged  equations and demonstrate their predictive capability.
 



%
%

The structure of the rest of the  paper is as follows. In section~\ref{sec2}, we first introduce the governing equations for phase field, mass and momentum transport along with boundary conditions  for incompressible, immiscible binary-fluids at the pore scale.
In Section~\ref{sec3}, the essential elements of the volume averaging method are briefly recalled, and the upscaled equations are derived based on spatial and temporal averaging theorems.
In Section~\ref{sec4},  we applied the length-scale constraints and necessary assumptions to  the derived equations at the Darcy scale, and the closed average model is built by addressing the corresponding closure problems.
In Section~\ref{sec5}, some numerical tests will be simulated to evaluate the capability of the present model.
Finally, in Section~\ref{sec6}, we summarize our conclusions and give some remarks for future research.

\section{Pore-scale equations}~\label{sec2}
We consider the incompressible and Newtonian flow of two-phase fluids completely filling a rigid  porous media.  The solid phase is assumed to be rigid and impermeable.  The entire pore space is denoted by $\Omega$ and its boundary is denoted by $\partial\Omega$.
In the diffuse-interface theory, the interface has a small but finite thickness, inside which the two fluids can be mixed and store a mixing energy.
Thus, an order parameter $\phi$ that smoothly varies in $[0,1]$ is introduced to represent the two fluids, i.e. the wetting fluid is labeled by $\phi=1$,  the non-wetting fluid is marked by $\phi=0$, and the fluid-fluid interface is marked by $\phi=0.5$. The system free-energy function dependent on the order parameter and its spatial derivative can be written as
\begin{equation}\label{eq}
F(\phi,\nabla\phi)=\int_\Omega\left[f(\phi)+\frac{\kappa}{2}|\nabla\phi|^2\right] d\Omega+ \int_{\partial\Omega} f_w(\phi) d\partial\Omega,
\end{equation}
where $f(\phi)=\lambda\phi^2(1-\phi)^2$ is the bulk free-energy density representing phase separation,   $\lambda$ is the mixing energy density,
the second term in the brackets describes the interfacial energy, $\kappa$ is the interfacial energy density, the last term $f_w$ is the wall free energy  related to the wetting behavior.
\textcolor{red}{Based on the formulations of the wall free energy, many wetting boundary conditions have been developed, including 
Linear form and cubic form~\cite{huang2015wetting}.  Compared to the linear form, the cubic form can avoid the appearance of a layer of spurious film at the solid surface. Hence, the cubic form is employed in this study~\cite{jacqmin2000contact},}
\begin{equation}\label{eq}
  f_w=-\sigma \cos(\theta)(\phi-0.5)(-2\phi^2+2\phi+1)+\frac{\sigma_{s,w}+\sigma_{s,nw}}{2},
\end{equation}
where $\theta$ is the static contact angle, $\sigma$ is  surface tension coefficient between the wetting and non-wetting fluids, $\sigma_{s,w}$ and $\sigma_{s,nw}$ are the fluid-solid interfacial tension for the wetting fluid and non-wetting fluid, respectively. Based on Young's equation, $\sigma_{s,w}-\sigma_{s,nw}=\sigma\cos\theta$.

The variation of the  free-energy \textcolor{red}{functional} with respect to the order parameter yields the following chemical potential
 \begin{equation}\label{eq:pore-chemical}
\mu=\lambda(4\phi^3-6\phi^2+2\phi)-\nabla \cdot \kappa \nabla\phi,
 \end{equation}
 and the boundary condition accounting for the wetting behavior
 \begin{equation}\label{eq:pore_wettingboundary}
\kappa \nabla\phi\cdot\bm n =-\frac{\partial f_w(\phi) }{\partial\phi},\qquad  \text{on}  \quad \partial\Omega
 \end{equation}
where  $\bm n$ denotes the unit vector normal to the solid wall $\partial\Omega$ pointing out of $\Omega$.

\textcolor{red}{
When considering a one-dimensional planar interface at equlibirum, the chemical potential should be uniform and minimum. Based on Eq.(\ref{eq:pore-chemical}), we can obtain the equilibrium profile of the order parameter
\begin{equation}
\phi(x)=\frac{1}{2}+\frac{1}{2}\tanh\left(\sqrt{\frac{\lambda}{2\kappa}}x\right),
\end{equation}
where $\sqrt{\lambda/2\kappa}$ is related to the interface thickness. 
If letting $2/W= \sqrt{\lambda/2\kappa}$, one can otain that the interface thickness W that be the distance from $\phi=0.018$ to $\phi=0.982$ contains $96.4\%$ of the variation of the order parameter. Furthermore, we can calculate the surface tension defined as the energy per unit surface area, 
\begin{equation}\label{sigma}
\sigma=\int \left(f(\phi)+\frac{\kappa}{2}|\partial_x\phi|^2 \right)dx=\frac{\sqrt{2\lambda \kappa}}{6}.
\end{equation}
To facilitate the simulation, $\lambda$ and $\kappa$ are widely calculated by setting the interfacial thickness $W$ and the surface tension coefficient, i.e,  $\lambda=12\sigma/W$ and $\kappa=3\sigma W/2$.}

The transport of two fluids at the pore scale can be formulated as the
  Navier-Stokes and Cahn-Hilliard equations
\begin{align}
\label{eq:CH}
\partial_t\phi+\nabla\cdot(\phi \bm u)&=\nabla\cdot M\nabla\mu,\\
\label{eq:NS}
\partial_t(\rho\bm u)+\nabla\cdot(\rho \bm u\otimes\bm u)&=-\nabla p+\nabla\cdot(2\eta  \bm{D(\bm u)}) +\bm F_{sf} +\rho \bm g,\\
\label{eq:div_u}
\nabla\cdot \bm u&=0, \\
\rho &=(\rho_1-\rho_2)\phi+\rho_2, \\
\eta &=(\eta_1-\eta_2)\phi+\eta_2,
\end{align}
where $ p$ is the pressure, $\eta $ is the fluid viscosity,
$\bm{D}(\bm u)=(\nabla\bm u+\nabla \bm u^T)/2$ is the symmetric velocity gradient tensor, $\bm F_{sf}=-\phi\nabla\mu$ is the interfacial force, 
$\bm g$ is the gravitational acceleration.
 $\rho$ is the density, the subscripts $1$ and $2$ denote different fluids,  $M$ is the mobility coefficient,
 $\mu$ is the chemical potential. Eq.(\ref{eq:CH}) is known as Cahn-Hillard equation designed to capture the evolution of the interface between two phases.
No-slip boundary for the velocity and the Neumann boundary condition for the chemical potential  $\mu$ read
\begin{equation}\label{eq:boundary}
\begin{aligned}
 \bm u &=0, \qquad \text{on} \quad \partial\Omega \\
\nabla\mu\cdot \bm n &=0,\qquad  \text{on}  \quad\partial\Omega \\
\end{aligned}
\end{equation}
where $\bm n$ is the normal vector to wall pointing out of the fluid. The Neumann boundary condition for the chemical potential implies the mass-conservation law.

\section{Derivation of the upscaled  equations}~\label{sec3}
\subsection{Overview of the volume average method}
\begin{figure}
  \centering
  \includegraphics[width=0.8\textwidth]{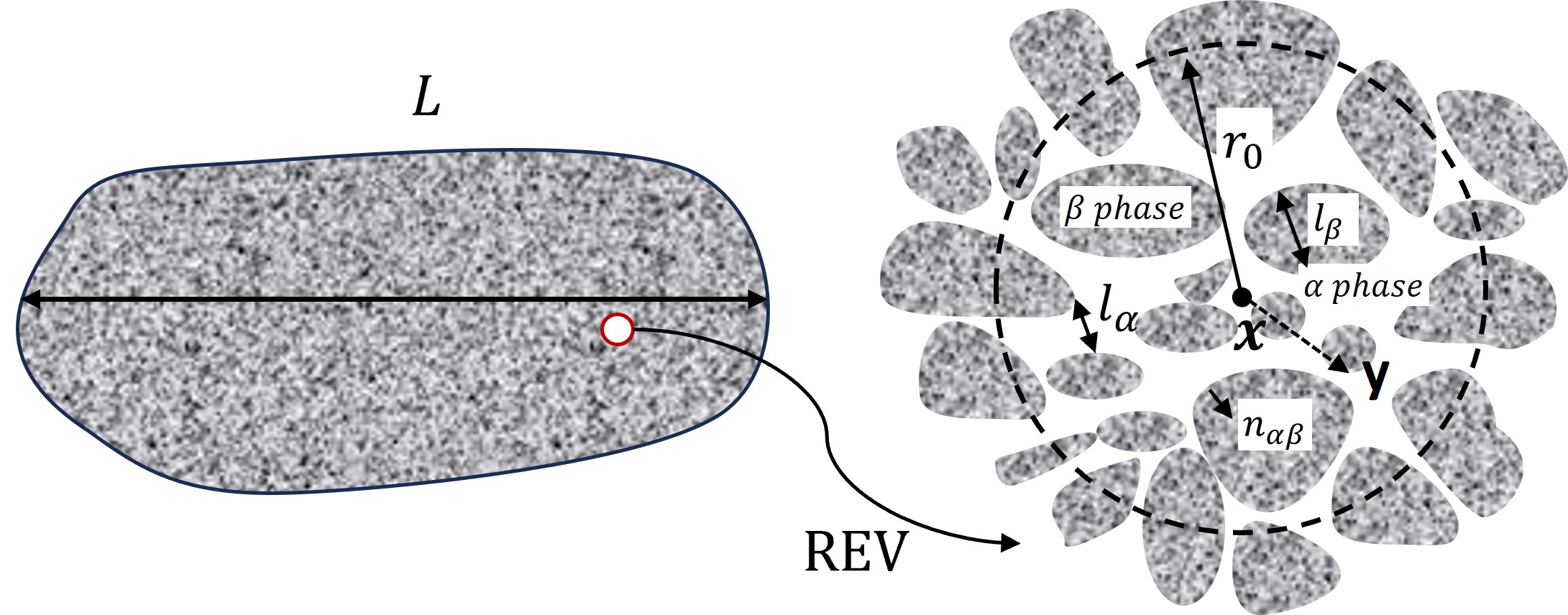}
  \caption{Schematic representation of the hierarchy of length scales of a model porous medium and of a typical representative volume}\label{fig:skeptch}
\end{figure}
In order to upscale the pore-scale governing equations using the volume averaging method, we first define an averaging domain $V$ that contains
both solid  $\beta$,  fluid phase $\alpha$, and \textcolor{red}{ $V_\alpha+V_\beta=V$}. 
\textcolor{red}{Without loss of generality, we consider the pore space filled with two immiscible fluids labeled as the wetting phase $\phi_w $ and the non-wetting phase $\phi_{nw}$}.
 As sketched in Fig.~(\ref{fig:skeptch}),   some characteristic lengths have been specified, i.e, the characteristic length of fluids  occupying the pores $l_{\alpha}$, the characteristic length of solid $l_{\beta}$, the characteristic length of the pore size $l=max(l_{\beta},l_{\alpha})$,  the characteristic length of the averaging volume $r_0$, and the characteristic length of the macroscopic domain $L$.
 To filter out unnecessary information at the macroscopic level,  length-scale constraints  are necessary. Following the  postulates in ~\cite{howes1985spatial,wood2013volume}, the length-scale should satisfy
\begin{equation}\label{eq:length-scale}
l\ll r_0  \ll L.
\end{equation}
As a result, the averaging volume $V$ can be regarded as  representative elementary volume (REV).
Besides, there is a disparity between the corresponding characteristic time scales, and these time scales should satisfy the following constraint
\begin{equation}\label{eq}
t_{l} \ll t_{r_0} \ll t_{L},
\end{equation}
where $t_l$ denotes the transport time of a certain quantity of fluid at the pore scale, $r_0$ denotes  the transport time of the corresponding average quantity at the REV, $L$ indicates the macroscopic transport time.

We define $\psi_\alpha$ as the value of the phase indicator $\psi$ in the $\alpha$-$\text{phase}$, such as density and velocity, and its value is zero in the  $\beta$-$\text{phase}$.
In terms of this averaging domain,
the superficial average of  $\psi_\alpha$   is introduced for a continuous smooth function  and defined  as
\begin{equation}\label{eq:superfical}
\langle\psi(\bm x)\rangle=\frac{1}{V}\int_{\bm y\in V_{\alpha}(\bm x)}\psi (\bm x+\bm y)d V,
\end{equation}
where  $\bm x$ represents the centroid of averaging volume, the vector $\bm y$ represents a displacement relative to the location $\bm x$; $V_{\alpha}=V_w+V_{nw}$ is the domain occupied by the fluids in the  averaging domain and is a continuous function of $\bm x$;  $\bm y\in V_{\alpha}(\bm x)$ indicates the set of all possible vectors such that displacements from the centroid yield points that are within the averaging domain $V_{\alpha}(\bm x)$. \textcolor{red}{It is noted that the subscripts $nw$ and $w$   denote the non-wetting fluid and wetting fluid, respectively.}
In contrast to the superficial average, the intrinsic average  of  $\psi_{\alpha}$    is formed by integrating only over the $\alpha$-phase,
\begin{equation}\label{eq:intrinsic}
\langle\psi_{\alpha}(\bm x)\rangle^{\alpha}=\frac{1}{V_{\alpha}}\int_{\bm y\in V_{\alpha}(\bm x)}\psi_{\alpha}(\bm x+\bm y) dV ,
\end{equation}
 For notational convenience, these definitions are annotated as
\begin{equation}\label{eq:superfical_simplify}
\langle\psi_{\alpha}\rangle=\frac{1}{V}\int_{V_{\alpha}}\psi_{\alpha} d V,  \qquad
\langle\psi_{\alpha}\rangle^{\alpha}=\frac{1}{V_{\alpha}}\int_{V_{\alpha}}\psi_{\alpha} d V.
\end{equation}
The two averaging operators are related by $\langle\psi_{\alpha}\rangle=\epsilon\langle\psi_{\alpha}\rangle^{\alpha}$, where $\epsilon=V_{\alpha}/V$ is the porosity in REV,
In order to obtain the macro-scale equations in terms of intrinsic phase averages from the pore-scale equations,
 it is necessary to interchange spatial differentiation and integration, which can be performed by the following spatial and temporal averaging theorems~\cite{howes1985spatial,whitaker2013method}
\begin{subequations}\label{eq:averaging_Theorems}
\begin{align}
\label{eq:averaging_Theorems1}
\langle\frac{\partial \psi_{\alpha}}{\partial t}\rangle= \frac{\partial \langle\psi_{\alpha}\rangle}{\partial t}- \frac{1}{V}\int_{A_{\alpha\beta}} \psi_\alpha \bm{\omega}_{\alpha\beta} \cdot \bm n_{\alpha\beta} dA\\
\label{eq:averaging_Theorems2}
\langle\nabla\psi_{\alpha}\rangle=\nabla \langle \psi_{\alpha}\rangle+\frac{1}{V}\int_{A_{\alpha\beta}}\psi_{\alpha} \bm n_{\alpha\beta} dA,
\end{align}
\end{subequations}
where $A_{\alpha\beta}=A_{w\beta}+A_{nw\beta}$ denotes the area of the fluid-solid interface within the averaging volume, $\bm n_{\alpha\beta}$ is a unit normal vector along the $\alpha$-$\beta$ interface pointing outward from the $\alpha$ phase,   $\bm{\omega}_{\alpha\beta}$ is the velocity of the $\alpha$-$\beta$ interfacial surface.
Here we assume that no-slip velocity boundary condition holds on the surface of the solid grains.
Especially, applying the spatial averaging theorem Eq.~(\ref{eq:averaging_Theorems}) to $\psi_{\alpha}=1$ leads to the very useful lemma
\begin{equation}\label{eq:psi=1}
\begin{aligned}
  \nabla\epsilon=-\frac{1}{V}\int_{A_{\alpha\beta}}\bm n_{\alpha\beta}dA.
\end{aligned}
\end{equation}

A key element in the volume averaging method is the Gray's decomposition~\cite{gray1975derivation}, namely,  the pore-scale quantities can be decomposed into their  intrinsic averages and their deviations
\begin{equation}\label{eq:decomposition}
\psi_\alpha=\langle\psi_\alpha \rangle^{\alpha}+\hat{\psi}_\alpha.
\end{equation}
Physically,  the average quantities are slow-varying fields and the spatial deviations are fast-varying fields.
Inserting Eq.~(\ref{eq:decomposition}) into Eq.~(\ref{eq:averaging_Theorems2})  and also applying Eq.(\ref{eq:psi=1}) lead to
\begin{equation}\label{eq:modify_averaging}
\begin{aligned}
\langle\nabla \psi_\alpha \rangle=\epsilon \nabla \langle \psi_\alpha \rangle^{\alpha} +\frac{1}{V}\int_{A_{\alpha\beta}}\hat{\psi}_\alpha \bm n_{\alpha\beta} dA,
\end{aligned}
\end{equation}
It is noted that the quantity $\psi$ is defined for every point in the considered volume, regardless of the phase. Hence,  one can have $\psi$, $\langle \psi\rangle^{\alpha}$ and $\hat{\psi}$ for any point $\bm x$.

To proceed further with the analysis, the following  assumptions are made :
\begin{enumerate}
\item The porosity is assumed to  vary only in space, then, $\partial_t \epsilon=0$.
\item The volume-averaged quantities are well-behaved, and $\langle\hat{\psi}_\alpha\rangle=0$ for any quantity.
\item Local equilibrium  holds in the REV,  that is,  the chemical potential is constant in each REV, but the averaged chemical pontial in different REV may be different,  that is $\langle\mu \rangle^\alpha=const$,  $\nabla \langle\mu \rangle^\alpha\neq 0$
\end{enumerate}
\textcolor{red}{It is worth pointing out that the local equilibrium hypothesis is justified as we are interested in the macroscopic transport.
It is obvious to consider that the time scale which corresponds to the evolution of the interface is faster than the one at the macroscopc scale.
Although the system is globally out of equilibrium,  equilibrium within each REV is established much faster than macroscopic transport processes. Consequently, the chemical potential can be treated as a spatial constant inside an individual REV, despite non-zero macroscopic gradients between adjacent REVs.} 
 In the following subsections, the coupled Navier-Stokes-Cahn-Hilliard equations will be averaged using the   volume averaging method and aforementioned assumptions.

\subsection{Volume averaging of Cahn-Hilliard equation}
In view of the Cahn-Hilliard equation, we average Eq.(\ref{eq:CH}) as
\begin{equation}\label{eq:av_CH}
\langle \partial_t \phi \rangle +
\langle \nabla \cdot(\phi \bm u) \rangle=\langle \nabla \cdot M\nabla \mu  \rangle ,\\
\end{equation}
Application of the transport theorem Eq.~(\ref{eq:averaging_Theorems1}) to the time term of Eq.(\ref{eq:CH}), and considering no-slip boundary condition, one gets
\begin{equation}\label{eq:1_left_CH}
\langle\frac{\partial \phi}{\partial t} \rangle
=\frac{\partial}{\partial t}\langle \phi \rangle.
\end{equation}
Applying the transport theorem Eq.~(\ref{eq:averaging_Theorems2}) to the convective term, using the Gray's spatial decomposition and no-slip boundary condition, we have
\begin{equation}\label{eq:2_left_CH}
\begin{aligned}
\langle\nabla\cdot(\phi \bm u)\rangle=&\nabla \cdot \langle(\langle \phi\rangle^{\alpha}+\hat{\phi})
(\langle \bm u\rangle^{\alpha}+\hat{\bm u})  \rangle\\
&=\nabla\cdot  (\epsilon \langle \phi\rangle^{\alpha}\langle \bm u\rangle^{\alpha})
+\nabla\cdot  \langle \hat{\phi}\hat{\bm u}\rangle.
\end{aligned}
\end{equation}
Similarly, the same operation for the diffusion term on the right hand side of  Eq.(\ref{eq:CH}) leads to
\begin{equation}\label{eq:1_right_CH}
\begin{aligned}
\langle \nabla \cdot (M\nabla \mu) \rangle &=
\nabla\cdot \langle M\nabla \mu \rangle+\frac{1}{V}\int_{A_{\alpha\beta}}M\nabla\mu\cdot \bm n_{\alpha\beta} dA \\
&=\nabla\cdot(M\epsilon \nabla\langle\mu \rangle^{\alpha} )
\end{aligned}
\end{equation}
where Eq.(\ref{eq:boundary}) and the local thermal equilibrium hypothesis are used.  Inserting all terms as derived above into Eq.(\ref{eq:av_CH}), we have
\begin{equation}\label{eq:VA_CH}
\begin{aligned}
\frac{\partial}{\partial t}\epsilon\langle  \phi\rangle^{\alpha}
+\nabla\cdot (\epsilon\langle \phi \rangle^{\alpha}\langle \bm u \rangle^{\alpha})
=
\nabla\cdot (M\epsilon\nabla \langle\mu \rangle^{\alpha})-\nabla\cdot \langle \hat{\phi}\hat{\bm u} \rangle.
\end{aligned}
\end{equation}
Note that the deviation term $\langle \hat{\phi}\hat{\bm u} \rangle $ is  unknown and will be determined in the next section.
As we focus on the formulation of the capillary pressure at the Darcy scale that is closely related to the pore-scale force $F_{sf}=-\phi\nabla \mu$,
the averaged chemical potential is very important.  As the chemical potential at the pore scale is well defined, application of the transport theorems to Eq.(\ref{eq:pore-chemical}) leads to the averaged chemcial potential
\begin{equation}\label{eq:VA_Mu}
\begin{aligned}
\langle\mu_{\phi}\rangle^\alpha &=
-\kappa  \nabla^2\langle \phi \rangle^\alpha
-\frac{\kappa}{\epsilon}\nabla \langle \phi \rangle^\alpha \cdot  \nabla\epsilon
-\frac{\kappa}{\epsilon}\nabla\cdot\frac{1}{V}\int_{A_{\alpha\beta}} \hat{\phi}\bm n_{\alpha\beta}dA
- \frac{1}{\epsilon}6\sigma \cos(\theta) a_v(\langle \phi\rangle_A-\langle\phi^2\rangle_A)
\\
&
+4\lambda (\langle \phi \rangle^{\alpha})^3
 + 12\lambda \langle (\hat{\phi})^2  \rangle^\alpha \langle \phi \rangle^{\alpha}
 + 4\lambda \langle (\hat{\phi})^3  \rangle^{\alpha}
    -6\lambda (\langle \phi \rangle^{\alpha})^2
 -6\lambda \langle (\hat{\phi})^2  \rangle^{\alpha}+2\lambda \langle \phi \rangle^{\alpha}
\end{aligned}
\end{equation}
where the wetting boundary condition has been applied, $a_v=A_{\alpha\beta}/V$ is the surface  area per unit volume,  and $\langle \phi \rangle_A$ is an area-averaged order parameter defined by
\begin{equation}\label{eq}
  \langle \phi \rangle_A=\frac{1}{A_{\alpha\beta}}\int_{A_{\alpha\beta}} \phi dA
\end{equation}
The presence of $\langle\hat{\phi}\rangle$ introduces unclosed terms. A closure model must be developed explicitly to complete the averaged chemical potential.


\subsection{Volume averaging of the continuity equation}
The superficial average of the continuity equation becomes
\begin{equation}\label{eq:continue_va1}
\langle\nabla\cdot\bm u\rangle=\nabla\cdot\langle\bm u\rangle +\frac{1}{V}\int_{A_{\alpha\beta}} \bm u\cdot\bm n_{\alpha\beta} dA=0,
\end{equation}
Using the no-slip boundary condition leads to the averaged continuity equation
\begin{equation}\label{eq:continue_va2}
  \nabla\cdot\langle\bm u\rangle=0.
\end{equation}
The above equation can also be rewritten as
\begin{equation}\label{eq:continue_va3}
\nabla\cdot\langle\bm u\rangle^{\alpha}=-\langle\bm u \rangle^{\alpha} \cdot \frac{\nabla\epsilon}{\epsilon}.
\end{equation}
Subtraction of Eq.~(\ref{eq:continue_va2}) from Eq.~(\ref{eq:continue_va3}) provides the spatial deviation equation for the velocity
\begin{equation}\label{eq:continue_va4}
\nabla\cdot\hat{\bm u}=\langle\bm u \rangle^{\alpha}\cdot  \frac{\nabla\epsilon}{\epsilon}.
\end{equation}
where the spatial decomposition and the length-scale restrictions are used. Eq.(\ref{eq:continue_va4}) implies that the divergence of the deviation velocity $\langle \hat{\bm {u}}\rangle$ has a non-zero value when the porosity varies.

\subsection{Volume averaging of the momentum equation}
The superficial average of the momentum equation becomes
\begin{equation}\label{eq:VA_NS}
\langle \frac{\partial \rho \bm u}{\partial t} \rangle
+\langle \nabla\cdot (\rho \bm u\bm u) \rangle
=- \langle \nabla p \rangle
+\langle\nabla\cdot \eta(\nabla\bm u+(\nabla\bm u)^T) \rangle
-\langle \phi\nabla\mu \rangle
+\langle \rho \bm{g} \rangle.
\end{equation}
Applying the averaging theorem to the first term of Eq.(\ref{eq:VA_NS}), we have
\begin{equation}\label{eq:1_left_NS}
\begin{aligned}
\langle\frac{\partial\rho \bm u}{\partial t}\rangle
&=\frac{\partial \langle \rho\bm u\rangle}{\partial t}
-\frac{1}{V}\int_{A_{\alpha\beta}}\rho\bm u \bm{\omega}_{\alpha\beta} \cdot \bm n_{\alpha\beta}dA
\\
&=\frac{\partial \epsilon \langle \rho\rangle^{\alpha} \langle\bm u\rangle^{\alpha}}{\partial t}
+\frac{\partial \langle\hat{\rho}\hat{\bm u}\rangle}{\partial t},
\end{aligned}
\end{equation}
Applying the averaging theorem Eq.~(\ref{eq:averaging_Theorems}) to the second term  of Eq.(\ref{eq:VA_NS}), we have
\begin{equation}\label{eq:2_left_NS}
\begin{aligned}
\langle \nabla\cdot(\rho\bm u\bm u)\rangle&=
\nabla\cdot\langle\rho \bm u\bm u\rangle
+\frac{1}{V}\int_{A_{\alpha\beta}} \rho\bm u\bm u\cdot\bm n_{\alpha\beta}dA
\\
=&\nabla\cdot(\epsilon \langle\rho\rangle^{\alpha} \langle\bm u\rangle^{\alpha} \langle\bm u\rangle^{\alpha} )
+
\nabla\cdot(\langle\rho\rangle^{\alpha} \langle\hat{\bm u}\hat{\bm u}\rangle )
\\
&+\nabla\cdot(\langle\bm u\rangle^{\alpha}
 \langle \hat{\rho}\hat{\bm u}\rangle)
+
\nabla\cdot( \langle \hat\rho\hat{\bm u}\rangle\langle\bm u\rangle^{\alpha})
+
\nabla \cdot \langle \hat{\rho}\hat{\bm u}\hat{\bm u}\rangle,
\end{aligned}
\end{equation}
Similarly, the average pressure term can be treated as
\begin{equation}\label{eq:1_right_NS}
\begin{aligned}
\langle\nabla p\rangle &=  \nabla \epsilon \langle p \rangle^{\alpha}
+\frac{1}{V}\int_{A_{\alpha\beta}} p \bm n_{\alpha\beta} dA,\\
&=\epsilon \nabla\langle p\rangle^{\alpha}+\frac{1}{V}\int_{A_{\alpha\beta}}\hat{p} \bm n_{\alpha\beta}dA.
\end{aligned}
\end{equation}
The averaging of the viscous stress term reads
\begin{equation}\label{eq:2_right_NS}
\begin{aligned}
\langle\nabla\cdot 2\eta \bm D(\bm u)\rangle
&= \nabla\cdot\langle \eta(\nabla \bm u+\nabla \bm u ^T)\rangle
+\frac{1}{V}\int_{A_{\alpha\beta}}\eta(\nabla \bm u+\nabla \bm u^T)\cdot \bm n_{\alpha\beta} dA, \\
=
&\nabla\cdot \left[ \epsilon \langle \eta \rangle^{\alpha} (\nabla\langle\bm u\rangle^{\alpha}+ (\nabla\langle\bm u\rangle^{\alpha})^T)
+
\langle \eta (\nabla\hat{\bm u}+(\nabla\hat{\bm u})^T) \rangle \right]
+
\frac{1}{V}\int_{A_{\alpha\beta}}\eta(\nabla \bm u+ (\nabla \bm u)^T)\cdot \bm n_{\alpha\beta} dA \\
=&
\nabla\cdot\left(  \langle \eta\rangle^{\alpha} (\nabla \epsilon\langle \bm u \rangle^{\alpha}+(\nabla \epsilon \langle\bm u\rangle^{\alpha})^T)\right)
- \langle\eta \rangle^\alpha\left[\nabla\langle \bm u\rangle^\alpha+(\nabla\langle \bm u\rangle^\alpha)^T  \right]\cdot \nabla\epsilon
\\
&
+\nabla\cdot(\langle \hat{\eta} (\nabla\hat{\bm u}+(\nabla\hat{\bm u} )^T) \rangle)
+ (\nabla\langle\bm u\rangle^{\alpha} +(\nabla\langle \bm u \rangle^{\alpha})^T)\cdot \langle \nabla\hat{\eta}\rangle
+\frac{1}{V}\int_{A_{\alpha\beta}} \eta (\nabla \hat{\bm u}+(\nabla \hat{\bm u})^T)\cdot\bm n_{\alpha\beta}dA.
\end{aligned}
\end{equation}
where \textcolor{red}{the no-slip  boundary is enforced at the solid surface. Specifically, the velocity and viscoisty are decomposed as , $\hat{\bm u}=-\langle\bm u\rangle^\alpha$ and $\eta=\langle\eta\rangle^\alpha+\hat{\eta}$.}
With the local equilibrium assumption, 
the averaging of the interfacial force term becomes
\begin{equation}\label{eq:3_right_NS}
\begin{aligned}
\langle\phi\nabla\mu\rangle=\epsilon \langle\phi \rangle^{\alpha} \nabla\langle \mu \rangle^{\alpha}.
\end{aligned}
\end{equation}
The averaging of the gravity force term reads
\begin{equation}\label{eq:4_right_NS}
\begin{aligned}
\langle \rho \bm{g}\rangle &=\epsilon \langle \rho\rangle^{\alpha}\bm{g},
\end{aligned}
\end{equation}
where $\langle\rho \rangle^{\alpha}=(\rho_1-\rho_2)\langle \phi \rangle^{\alpha}+\rho_2$.

Substitution of Eqs.(\ref{eq:1_left_NS})-(\ref{eq:4_right_NS}) into Eq.(\ref{eq:VA_NS}),    the volume averaged momentum equation  becomes
\begin{equation}\label{eq:vA_NS_final}
\begin{aligned}
\frac{\partial(\epsilon\langle\rho\rangle^{\alpha}\langle\bm u\rangle^{\alpha})}{\partial t}
+
\nabla \cdot(\epsilon\langle\rho \rangle^{\alpha} \langle\bm u \rangle^{\alpha} \langle\bm u \rangle^{\alpha} )
&=-\epsilon \nabla \langle p\rangle^{\alpha}
+
 \nabla\cdot \langle\eta\rangle^{\alpha} \left( \nabla\epsilon\langle\bm u \rangle^{\alpha}+(\nabla \epsilon\langle \bm u\rangle^{\alpha})^T \right) \\
&- \langle\eta \rangle^\alpha\left[\nabla\langle \bm u\rangle^\alpha+(\nabla\langle \bm u\rangle^\alpha)^T  \right]\cdot \nabla\epsilon
-\epsilon \langle\phi \rangle^{\alpha} \nabla\langle \mu \rangle^{\alpha} +\epsilon \langle\rho\rangle^\alpha \bm{g} \\
&
+\hat{\bm R}_{t}
+\hat{\bm R}_{u}
+\hat{\bm R}_{\eta}+\hat{\bm R}_{drag}
\end{aligned}
\end{equation}
with
\begin{equation}\label{eq:R}
\begin{aligned}
\hat{\bm R}_{t} =&-\frac{\partial\langle \hat{\rho} \hat{\bm u}\rangle }{\partial t}, \\
\hat{\bm R}_{u} =&-\nabla\cdot(\rho \langle\hat{\bm u}\hat{\bm u} \rangle
+\langle \bm u\rangle^{\alpha}\langle\hat{\rho}\hat{\bm u} \rangle
+\langle\hat{\rho}\hat{\bm u} \rangle\langle\bm u \rangle^{\alpha}
), \\
\hat{\bm R}_{\eta} =& \nabla\cdot(\langle \hat{\eta} (\nabla\hat{\bm u}+\nabla\hat{\bm u}^T)\rangle)
+\langle \nabla\hat{\eta}\rangle \cdot(\nabla\langle\bm u \rangle^{\alpha}+(\nabla\langle\bm u\rangle^{\alpha})^T)
\\
\bm{\hat R}_{drag}=& \frac{1}{V}\int_{A_{\alpha\beta}}(-\hat{p}\bm I+\eta (\nabla\hat{\bm u}+ (\nabla \hat{\bm u})^T))\cdot\bm n_{\alpha\beta} dA.
\end{aligned}
\end{equation}
where $\hat{\bm R}_{t}, \hat{\bm R}_{u}$, $\hat{\bm R}_{\eta}$, $\hat{\bm R}_{drag}$ are unclosed terms due to the existences of $\hat{\rho}\hat{\bm u}$ and $\hat{\eta}$.
\textcolor{red}{
To this point, all averaged equations have been obtained, and 
all unclosed terms remains entirely without extra assumptions.
It can be observed that the formulations are similar to the pore scale ones. However, the averaged governing equations incorporate unclosed terms arising from 
volume averages  and area integrals of  spatial deviations.
These terms  can be treated as volume and surface filters~\cite{whitaker2013method}, respectively. These filtering operations 
are central to   the volume averaging method.
The microscale  information retained by these integrals will be resolved in the subsequent closure problem. }

\section{Closure of the two-phase Darcy scale equations}~\label{sec4}
\subsection{Closure of the averaged Cahn-Hilliard equation}
\textcolor{red}{
In previous studies, the averaged surface tension force is roughly treated as the 
capillary pressure, whose formulations have been widely stuidied by the core-scale flow experiments~\cite{chen2019homogenization}
. However, this kind of approximation inevitably neglects certain important microscopic flow information, i.e, the wetting effect.
To fully identify the averaged chemical potential characteristics and achieve closure of  the averaged Cahn-Hilliard equation, it is necessary to directly identify the solution of $\hat{\phi}$.}
The governing equation for $\hat{\phi}$ can be obtained by subtracting Eq.~(\ref{eq:VA_CH}) from Eq.~(\ref{eq:CH}),
\textcolor{red}{
\begin{equation}\label{eq:CH_spatial_deviation}
\begin{aligned}
\frac{\partial \hat{\phi}}{\partial t}+\bm u\cdot\nabla\hat{\phi}+\bm{\hat{u}}\cdot \nabla\langle \phi\rangle^\alpha
=
-M\nabla\langle \mu \rangle^{\alpha}\cdot \frac{\nabla\epsilon}{\epsilon}
+\frac{\nabla\epsilon}{\epsilon}\cdot \langle\hat{\phi}\hat{\bm u} \rangle^\alpha
+\nabla\cdot\langle \hat{\phi}\hat{\bm u} \rangle^\alpha
\end{aligned}
\end{equation}
Recalling the definitions of the characteristic sizes, the advection terms in Eq.~(\ref{eq:CH_spatial_deviation}) can be estimated by
\begin{equation}\label{eq:closure_CH_3}
\nabla\cdot \langle \hat{\phi}\hat{\bm u} \rangle^\alpha 
=O\left(\frac{\hat{\phi} \langle\bm u\rangle^\alpha}{r_0 } \right).
\end{equation}
\begin{equation}\label{eq:closure_CH_4}
\bm u\cdot \nabla\hat{\phi}=
O\left(\frac{\langle \bm u\rangle^\alpha \hat{\phi} }{l}  \right).
\end{equation}
\begin{equation}
\frac{\nabla\epsilon}{\epsilon}\cdot\langle \hat{\phi}\hat{\bm u}\rangle^\alpha
=O\left(\frac{1}{r_0}\hat{\phi}\langle\bm u\rangle^\alpha \right),
\end{equation}
}
where $\bm{\hat{u}}$ is assumed to have the same amplitude  of $\langle\bm u\rangle$ considering $\bm u=0$ at the wall. Using the length-scale constraint, we deduce that
\begin{equation}\label{eq:inequation}
\begin{aligned}
\frac{1}{\epsilon}\nabla\cdot \langle \hat{\phi}\hat{\bm u} \rangle \ll  \bm u \cdot \nabla\hat{\phi}.
\end{aligned}
\end{equation}
Finally, the closure equation for $\hat{\phi}$ takes the following form,
\begin{equation}\label{eq:deviation_mu}
\frac{\partial \hat{\phi}}{\partial t}+\bm u\cdot \nabla \hat{\phi}
+\bm{\hat{u}}\cdot \nabla \langle \phi \rangle^{\alpha}
=
 -M\nabla\langle\mu \rangle^{\alpha} \cdot \frac{\nabla\epsilon}{\epsilon}.
\end{equation}
Similarly, the \textcolor{red}{corresponding} boundary condition  can be given by
\begin{equation}\label{eq:mu_boundary}
\nabla \langle \mu \rangle^{\alpha}\cdot \bm n=0 , \qquad \text{on} \qquad \partial \Omega
\end{equation}
\begin{equation}\label{eq:wetting_boundary}
\nabla\hat{\phi}\cdot \bm n=\frac{\partial f_w(\langle \phi \rangle^{\alpha},\theta)}{\partial \langle \phi \rangle^{\alpha}} -\nabla\langle \phi \rangle^{\alpha}\cdot \bm n,  \qquad \text{on}  \qquad \partial \Omega
\end{equation}
It is noted that the first term on the right-side hand in Eq.(\ref{eq:wetting_boundary})
is related to the wetting boundary condition.
In order to obtain   the  closure of  Eq.~(\ref{eq:VA_CH}), one also needs a representation for the spatial deviation term $\langle \hat{\phi}\hat{\bm u}\rangle$ .   From Eq.(\ref{eq:deviation_mu}) and the boundary condition Eq.(\ref{eq:wetting_boundary}),  the solution of $\hat{\phi}$  depends on $\langle \mu \rangle^\alpha$ and  $\langle \phi\rangle^\alpha$, thus it is reasonable to assume the solution of $\hat{\phi}$ has the following form
\begin{equation}\label{eq:symplify_hat_phi}
\begin{aligned}
\hat{\phi} &=\bm a_{\phi}\cdot \nabla  \langle \mu \rangle^{\alpha}+\bm{b}_\phi\cdot \nabla  \langle \phi \rangle^{\alpha}+ c_\phi \langle \phi \rangle^\alpha,  \\
\end{aligned}
\end{equation}
where $\bm a_\phi$ and $\bm b_\phi$ are vectors,  $c_\phi$ is a scalar, and satisfy $\langle \bm a_\phi \rangle =\langle\bm b_\phi\rangle=\langle c_\phi \rangle=0$ since $\langle\hat{\phi}\rangle=0$. 
Given that $\epsilon$=const, the right-hand side term in Eq.(\ref{eq:deviation_mu}) vanishes. Consequently, 
$\bm a_\phi$ depends only on the heterogeneity of the porous media.
 
In order to generate an estimate of $\hat{\bm u}$, considering the no-slip condition provides $\hat{\bm u}=-\langle\bm u \rangle^\alpha $,  the velocity deviation can be  represented as~\cite{whitaker1986flow}
\begin{equation}\label{eq:velocity_deviation}
  \hat{\bm u}=\bm B\cdot\langle\bm u\rangle^\alpha.
\end{equation}
where $\bm B$ is a second-rank tensor, which can be estimated in the next section.
Substitution Eqs.~(\ref{eq:symplify_hat_phi}) and (\ref{eq:velocity_deviation})  into Eq.~(\ref{eq:VA_CH})  leads to
\begin{equation}\label{eq:VA_CH_simplify}
\begin{aligned}
\frac{\partial}{\partial t}\epsilon\langle  \phi \rangle^{\alpha}
+\nabla\cdot (\epsilon(\bm I +\langle c_\phi \bm B\rangle^\alpha) \cdot \langle  \phi\rangle^{\alpha} \langle \bm u \rangle^{\alpha})
=
\nabla\cdot ( \epsilon \bm{M}_{eff}\cdot \nabla\langle\mu \rangle^{\alpha} )-\nabla\cdot( \bm B_\phi  \cdot \nabla \langle  \phi\rangle^{\alpha}  ),
\end{aligned}
\end{equation}
where the generalized diffusion coefficients $\bm{M}_{eff}$  and $\bm B_\phi$ are defined as
\begin{equation}\label{eq}
  \bm{M}_{eff}=  (M\bm I-\langle \bm B\cdot \langle\bm u\rangle^\alpha  \bm a_\phi \rangle^\alpha) , \qquad \bm B_\phi=\langle  \bm B\cdot \langle\bm u\rangle^\alpha  \bm b_{\phi}\rangle
\end{equation}
where  the first term in parentheses is the effective diffusivity tensor, and the second term is the hydrodynamic dispersion tensor.
Note that there are some unclosed parameters in $\langle \mu \rangle^\alpha $, i.e,  $\langle (\hat{\phi})^2   \rangle^\alpha $,$\langle (\hat{\phi})^3  \rangle^\alpha $.
Using Eq.(\ref{eq:symplify_hat_phi}), we can reformulate these terms
as
\begin{equation}\label{eq}
\begin{aligned}
  \langle (\hat{\phi})^2  \rangle^\alpha &=\lambda_{21} \langle \phi\rangle^\alpha+\lambda_{22} (\langle \phi\rangle^\alpha)^2 +\lambda_{23},
  \\
  \langle (\hat{\phi})^3  \rangle^\alpha &= \lambda_{31} \langle \phi\rangle^\alpha+ \lambda_{32} (\langle \phi\rangle^\alpha)^2 
  +\lambda_{33} ( \langle \phi\rangle^\alpha)^3 +\lambda_{34}.
\end{aligned}
\end{equation}
where $\lambda_{ij}$ are undetermined parameters, which could be determined by the closure problem in REV. 
In addition, the area-averaged order parameter can be decomposed in the form
\begin{equation}\label{eq:area_averaged_decomposition}
  \langle \phi \rangle_A=\langle \phi \rangle^\alpha +\bar{\phi}_A
\end{equation}
One can estimate the terms  at  the surface of a porous media as~\cite{ryan1981theory,kim1987diffusion}
\begin{equation}\label{eq}
\langle \kappa  \bm n\cdot \nabla\phi\rangle_A =O\left(\bar{\kappa}\frac{\langle \phi \rangle^\alpha -\langle \phi \rangle_A}{l}\right),
 \qquad
   -\langle \frac{\partial f_w}{\partial \phi}\rangle_A=O\left(6 {\sigma} \cos\theta (\langle \phi \rangle_A-\langle \phi\rangle_A \langle\phi \rangle_A )\right)
\end{equation}
where $\bar{\kappa}$ is parameter in the REV scale. Considering the wetting boundary conditions and Eq.(\ref{eq:area_averaged_decomposition}),  one can obtain
\begin{equation}\label{eq:area_averaged}
  \frac{\bar{\phi}_A}{\langle \phi\rangle^\alpha}=O\left[\frac{6\cos\theta \frac{ {\sigma} l}{\bar{\kappa}}\left( \frac{\langle\phi\rangle_A^2}{\langle\phi\rangle^\alpha}-1\right)}{ 1+6\cos\theta \frac{{\sigma} l}{\bar{\kappa}} }  \right]
\end{equation}
Here $l$ represents the characteristic length for the fluids and can be thought of as the mean pore diameter for a porous catalyst. When the restriction ${\sigma} l/\bar{\kappa}\ll 1$ is satisfied, $\langle\phi \rangle_A$
can be replaced by $\langle \phi \rangle^\alpha$. 
Then, the averaged chemical potential can be reformulated as
\begin{equation}\label{eq:VA_Mu_close}
\begin{aligned}
\langle\mu \rangle^\alpha &=
-\frac{1}{\epsilon}\nabla\cdot (\bar{\bm{\kappa}}_1 \cdot \nabla\langle \phi \rangle^\alpha)
-\frac{1}{\epsilon}\nabla\cdot \bar{\bm{\kappa}}_2 \langle \phi \rangle^\alpha
-\frac{6\sigma a_v}{\epsilon}\cos(\theta) \langle \phi \rangle^\alpha (1- \langle \phi \rangle^\alpha)
\\
&
+4\bar{\lambda}_3 (\langle \phi \rangle^{\alpha})^3
    -6\bar{\lambda}_2 (\langle \phi \rangle^{\alpha})^2
+2\bar{\lambda}_1  \langle \phi \rangle^{\alpha}
+\bar{\lambda}_0
\end{aligned}
\end{equation}
where
\begin{equation}\label{eq}
\begin{aligned}
\bar{\bm{\kappa}}_1=\kappa \epsilon \left(\bm I+\frac{1}{V_\alpha}\int_{A_{\alpha\beta}}\bm n_{\alpha\beta}\bm b_\phi dA\right),\qquad
\bar{\bm{\kappa}}_2=\kappa\epsilon\frac{1}{V_\alpha}\int_{A_{\alpha\beta}} \bm n_{\alpha\beta}c_\phi dA ,\\
  \bar{\lambda}_3=\lambda(1+\lambda_{33}+3\lambda_{22}), \qquad
  \bar{\lambda}_2=\lambda(1-\frac{2}{3}\lambda_{32}+\lambda_{22}-2\lambda_{21}),\\
  \bar{\lambda}_1=\lambda(\lambda_{31}-3\lambda_{21}),\qquad
  \bar{\lambda}_0=\lambda(4\lambda_{34}+6\lambda_{23}) .
\end{aligned}
\end{equation}
Up to this point, we have obtained the formulations of the expression of $\hat{\phi}$ and $\langle \mu\rangle^\alpha$.
The remaining issure is how to determine these coefficients. 

\subsection{Closure of the averaged momentum  equation}
The integral terms  $\hat{\bm R}_{drag}$ represent an average drag force acting on the fluids from the porous media.
Inspired by the expression of the fluid-solid interaction forces for single-phase flow in porous media~\cite{breugem2006derivation,ahammad2017numerical,hsu1990thermal},
the term $\bm \hat{R}_{drag}$   can be approximated by the following empirical linear formulation
\begin{equation}\label{eq:VA_NS_symplify_1}
\begin{aligned}
\frac{1}{V}\int_{A_{\alpha\beta}}\left[-\hat{p}\bm I+\eta(\nabla\hat{\bm u}+ (\nabla \hat{\bm u})^T)\right]\cdot\bm n_{\alpha\beta} dA
=-\epsilon\left[\frac{\epsilon \langle \eta\rangle^\alpha }{KK_r}+C_f\langle \rho\rangle^\alpha \epsilon^2 \frac{|\langle\bm u\rangle^{\alpha}|}{\sqrt{KK_r}} \right]\langle \bm u \rangle^{\alpha}.
\end{aligned}
\end{equation}
where $K_r$ is the relative permeability defined as $K_r=\langle \phi\rangle^\alpha K_{r,w}+(1-\langle \phi \rangle^\alpha) K_{r,nw}$, $K_{r,w}$ and $K_{r,nw}$ are the wetting and non-wetting phase relative permeability. 
$C_f$ represents the nonlinear Forchheimer coefficient. The first term on the right hand of Eq.(\ref{eq:VA_NS_symplify_1}) is considered as Darcy term for low flow velocity, and the second term is the non-Darcy term for high velocities.
It should be noted that various interpolation schemes are available for calculating $K_r$, each potentially influencing interfacial dynamics~\cite{cueto2009phase}. A systematic comparative analysis of these schemes exceeds the scope of this paper. 

Many relative permeability models are available in literature~\cite{stone1970probability,delshad1989comparison}.
For example, based on Ergun's experimental relationship, the geometric function $C_F$ and the absolute permeability $K$ of the porous medium can be defined as
\begin{equation}\label{eq:ergun}
C_F=\frac{1.75}{\sqrt{150\epsilon^3}},
 \qquad  K=\frac{\epsilon^3 d_p^2}{150(1-\epsilon)^2},
\end{equation}
where $d_p$ is the solid particle diameter.

Further progress can be made by means of additional assumptions.
First, assume  both density and viscosity differences are small,
and $\rho\approx\langle \rho \rangle^\alpha$. Then $\langle\hat{\rho}\hat{\bm u}\rangle=0$. As a result, one can obtain
\begin{equation}\label{eq:simplify_Ru}
\bm {\hat{R}}_{u}=-\nabla\cdot(\langle\rho\rangle^\alpha \langle \hat{\bm u}\hat{\bm u}\rangle ).
\end{equation}
Second, based on the model proposed by Breugem and Rees~\cite{breugem2006derivation},  the following formulation is employed
\begin{equation}\label{eq:VA_NS_symplify_2}
\nabla\cdot \langle \rho\rangle^\alpha \langle \hat{\bm u}\hat{\bm u}\rangle=
- \nabla\cdot \langle \eta_d\rangle^\alpha (\nabla \epsilon\langle \bm u\rangle^\alpha +(\nabla  \epsilon \langle\bm u\rangle^\alpha)^T )
+\langle\eta_d\rangle^\alpha (\nabla  \langle \bm u\rangle^\alpha +(\nabla   \langle\bm u\rangle^\alpha)^T  )\cdot \nabla\epsilon,
\end{equation}
where $\langle \eta_d\rangle^\alpha $ is the sub-filter viscosity due to dispersion. The sub-filter viscosity depends on the structural properties of the porous medium and on the flow.

Finally, combining Eqs.(\ref{eq:VA_CH}) and (\ref{eq:VA_Mu_close}), we arrive at the derived  volume averaged  NSCH system for multiphase flow with small density and viscosity contrasts in porous media
\begin{subequations}\label{eq:NSCH_homogenous}
\begin{align}
\frac{\partial }{\partial t} \epsilon \langle \phi\rangle^{\alpha}
+\nabla\cdot (\epsilon (\bm I+\langle c_\phi \bm B \rangle^\alpha)\cdot \langle \phi\rangle^{\alpha}\langle \bm u\rangle^{\alpha})
&=\nabla \cdot( \epsilon \bm M_{eff}\cdot \nabla \langle \mu \rangle^{\alpha})
- \nabla \cdot ( \langle \bm B\cdot \langle\bm u\rangle^\alpha  \bm b_\phi   \rangle \cdot \nabla \langle \phi \rangle^\alpha)
\\
\nabla \cdot (\epsilon \langle \bm u \rangle^{\alpha}) &=0,
\\
\frac{\partial (\epsilon \langle \rho\rangle^{\alpha}\langle \bm u\rangle^{\alpha}) }{\partial t}
+\nabla \cdot (\epsilon \langle \rho \rangle^{\alpha}\langle \bm u \rangle^{\alpha} \langle \bm u \rangle^{\alpha})
=&
- \epsilon\nabla \langle p\rangle^{\alpha}
+\nabla \cdot (\langle\eta \rangle_{eff}^\alpha) \left[\nabla \epsilon \langle \bm u \rangle^{\alpha}+(\nabla \epsilon \langle \bm u\rangle^{\alpha})^T \right] \\
& \hspace{-1.2cm}+\epsilon\langle \rho \rangle^\alpha\bm g
-\epsilon \langle \phi \rangle^\alpha\nabla \langle \mu \rangle^\alpha
-  \langle\eta \rangle_{eff}^\alpha (\nabla \langle \bm u\rangle^\alpha+(\langle \nabla\bm u\rangle^\alpha)^T )\cdot \nabla \epsilon
+\bm F_{av}
\end{align}
\end{subequations}
with
\begin{equation}\label{eq:mu_homogenous}
\begin{aligned}
\bm F_{av}=
&-\epsilon\left[\frac{\epsilon \langle \eta \rangle^\alpha }{KK_r}+C_f\langle \rho\rangle^\alpha \epsilon^2\frac{|\langle\bm u\rangle^{\alpha}|}{\sqrt{KK_r}} \right]\langle \bm u \rangle^{\alpha}
\\
\langle\rho\rangle^{\alpha} &= (\rho_1-\rho_2)\langle \phi\rangle^{\alpha}+\rho_2
\\
\langle\eta\rangle^{\alpha} &= (\eta_1-\eta_2)\langle \phi\rangle^{\alpha}+\eta_2
\end{aligned}
\end{equation}
where $\langle\eta \rangle_{eff}^\alpha=\langle \eta\rangle^{\alpha}+\langle \eta_d\rangle^\alpha$ is the viscosity of fluid saturating the porous medium (or effective viscosity). \textcolor{blue}{Both experimental and theoretical investigations indicate that the ratio of the effective viscosity to the base fluid viscoisty could be greater than unity or less than unity, or could be taken as unity. Some formulas have been developed to describe the effective viscoisty~\cite{almalki2016investigations}.
For simplicity, the effective viscosity $\langle\eta \rangle_{eff}^\alpha=\langle\eta \rangle^\alpha/\epsilon$ is used~\cite{goyeau2003momentum}.
}
To nondimensionalize the above equations , the following dimensionless variables are used,
\begin{equation}\label{eq}
\bm x^*=\frac{\bm x}{L_{ref}}, \quad  \bm u^*=\frac{\bm u}{\langle\bm u\rangle^\alpha_{ref}},t^*=\frac{t}{t_{ref}},\quad \lambda^*=\frac{\lambda}{\lambda_{ref}},\quad \epsilon^*=\frac{\epsilon}{\epsilon_0},
\end{equation}
 where the primed quantities are dimensionless, $\lambda_{ref}$ is  the characteristic energy, $L_{ref}$ is  the characteristic length,  $\langle\bm u\rangle^\alpha_{ref}$ is the characteristic velocity,  $t_{ref}=L_{ref}/\langle\bm u\rangle^\alpha_{ref}$ is the characteristic time, $\epsilon_0$ is  the characteristic porosity accounting for the varying porosity. Substituting these variables into Eq.(\ref{eq:NSCH_homogenous}) and dropping the primes, we have
\begin{equation}\label{eq:NSCH_homogenous}
\begin{aligned}
\frac{\partial }{\partial t} \epsilon \langle \phi\rangle^{\alpha}
+\nabla\cdot (\epsilon (\bm I+\langle c_\phi \bm B \rangle^\alpha)\cdot \langle \phi\rangle^{\alpha}\langle \bm u\rangle^{\alpha})
&=\frac{1}{\mathrm{Pe}}\nabla \cdot( \epsilon \bm M_{eff}\cdot \nabla \langle \mu \rangle^{\alpha})
- \nabla \cdot ( \langle \bm B\cdot \langle\bm u\rangle^\alpha  \bm b_\phi   \rangle \cdot \nabla \langle \phi \rangle^\alpha)
\\
\nabla \cdot (\epsilon \langle \bm u \rangle^{\alpha}) &=0,
\\
\frac{\partial (\epsilon \langle \rho\rangle^{\alpha}\langle \bm u\rangle^{\alpha}) }{\partial t}
+\nabla \cdot (\epsilon \langle \rho \rangle^{\alpha}\langle \bm u \rangle^{\alpha} \langle \bm u \rangle^{\alpha})
=&
- \epsilon\nabla \langle p\rangle^{\alpha}
+\frac{1}{\mathrm{Re}}\nabla \cdot (\langle\eta \rangle_{eff}^\alpha) \left[\nabla \epsilon \langle \bm u \rangle^{\alpha}+(\nabla \epsilon \langle \bm u\rangle^{\alpha})^T \right] \\
& \hspace{-1cm}+\frac{1}{\mathrm{Fo}^2}\epsilon\langle \rho \rangle^\alpha\bm g
-\frac{1}{\mathrm{We}}\epsilon \langle \phi \rangle^\alpha\nabla \langle \mu \rangle^\alpha \\
&-\epsilon\left[\frac{\epsilon_0}{\mathrm{Re} \mathrm{Da}}\frac{\epsilon \langle \eta \rangle^\alpha }{KK_r}+\frac{\epsilon^2_0}{\mathrm{Da}}C_f\langle \rho\rangle^\alpha \epsilon^2\frac{|\langle\bm u\rangle^{\alpha}|}{\sqrt{KK_r}} \right]\langle \bm u \rangle^{\alpha} \\
&-  \frac{1}{\mathrm{Re}}\langle\eta \rangle_{eff}^\alpha (\nabla \langle \bm u\rangle^\alpha+(\langle \bm u\rangle^\alpha)^T )\cdot \nabla \epsilon
\end{aligned}
\end{equation}
with
\begin{equation}\label{eq}
\begin{aligned}
\langle\mu \rangle^\alpha=& 4\bar{\lambda}_3 (\langle \phi \rangle^{\alpha})^3  
    -6\bar{\lambda}_2 (\langle \phi \rangle^{\alpha})^2
+2\bar{\lambda}_1  \langle \phi \rangle^{\alpha}
+\bar{\lambda}_0 \\
& -\nabla\cdot \mathrm{Cn}_1^2\nabla \langle\phi \rangle^\alpha-\nabla\cdot \mathrm{Cn}_2\langle\phi\rangle^\alpha
-6\epsilon^{-1} A_\sigma \cos(\theta)\langle\phi \rangle^\alpha (1-\langle\phi \rangle^\alpha).
\end{aligned}
\end{equation}
where $\mathrm{Pe}= L_{ref}\langle\bm u\rangle^\alpha_{ref}/\lambda_{ref}M$ is the Peclet number,
 $\mathrm{Re}=\langle \rho  \rangle^\alpha_{ref} \langle \bm u  \rangle^\alpha_{ref} L_{ref}/\langle \eta  \rangle^\alpha_{ref}$ is the Reynolds number,
 $\mathrm{We}=\langle \rho  \rangle^\alpha_{ref} (\langle \bm u  \rangle^\alpha_{ref})^2/\lambda_{ref}$ is the Weber number,
 $\mathrm{Fo}=\langle \bm u  \rangle^\alpha_{ref}/\sqrt{g L_{ref}}$ is the Frounde number,
 $\mathrm{Da}=\sqrt{K}/L_{ref}$ is the Darcy number,
 $\mathrm{Cn}_1=\sqrt{\bar{\bm{\kappa}}_1/L_{ref}^2\lambda_{ref}}$ and $\mathrm{Cn}_2=\bar{\bm{\kappa}}_2/L_{ref}\lambda_{ref}$ are the Cahn numbers.
 $\mathrm{A_{\sigma}}=a_v\sigma/\lambda\epsilon_0$ is a new dimensionless length. As $\mathrm{A_{\sigma}}$ increases, the system becomes dominated by the wetting behavior.

\textcolor{blue}{We provide the following comments:}

\begin{remark}
For a single-phase fluid flow with constant density and viscosity, the volume averaged CH equation and the surface tension force can be omitted. Under these conditions,   the averaged NS equations reduce  to
\begin{equation}
\begin{aligned}
\frac{\partial (\epsilon \langle \rho\rangle^{\alpha}\langle \bm u\rangle^{\alpha}) }{\partial t}
+\nabla \cdot (\epsilon \langle \rho \rangle^{\alpha}\langle \bm u \rangle^{\alpha} \langle \bm u \rangle^{\alpha})
=&
- \epsilon\nabla \langle p\rangle^{\alpha}
+\nabla \cdot [\langle\eta \rangle_{eff}^\alpha \left(\nabla \langle \bm u \rangle+(\nabla  \langle \bm u\rangle )^T \right)] \\
& \hspace{-1.2cm}+\epsilon\langle \rho \rangle^\alpha\bm g
-  \langle\eta \rangle_{eff}^\alpha (\nabla \langle \bm u\rangle^\alpha+(\langle \nabla \bm u\rangle^\alpha)^T )\cdot \nabla \epsilon
+\bm F_{av}.
\end{aligned}
\end{equation}
Our model bears a strong resemblance to
the formulation proposed by Nithiarasu~\cite{nithiarasu1997natural}. In Nithiarasu's model,  the pressure term is defined as $\nabla\epsilon\langle p\rangle^\alpha $ and 
the term $\langle\eta \rangle_{eff}^\alpha (\nabla \langle \bm u\rangle^\alpha+(\langle \nabla \bm u\rangle^\alpha)^T )\cdot \nabla \epsilon$ is omitted . Furthermore, our model is comparable to Gidaspow's model~\cite{gidaspow1994multiphase} with the primary difference residing in the  viscous term. Under the condition of spatially uniform porosity, these three models become identical.
\end{remark}

\begin{remark}
\textcolor{blue}{
At the pore scale, the viscous term in the incompressible NS equations with constant viscosity can be expressed as $\eta \nabla^2\bm u$ for a Newtonian fluid.
Applying the local volumetric average yields
\begin{equation}
\langle\eta\nabla^2\bm u \rangle =\eta\nabla^2\langle \bm u \rangle+\frac{1}{V}\int_{A_{\alpha\beta}}(\nabla\bm u)\cdot \bm n_{\alpha\beta}dA
=\eta\nabla\langle \bm u\rangle-\eta\nabla\langle\bm u\rangle^\alpha\cdot \nabla\epsilon+\frac{1}{V}\int_{A_{\alpha\beta}}(\nabla \hat{\bm u})\cdot \bm n_{\alpha\beta}dA.
\end{equation}
This is consistent with the model presented in Ref~\cite{whitaker1986flow}. In the present work, the viscous  term is taken in the form $\nabla\cdot \eta(\nabla \bm u+\nabla \bm u^T)$, which is equivalent to $\eta \nabla^2\bm u$ but retains a more general stress-tensor structure. Applying the volume average operator to this expression gives
\begin{equation}\label{full_viscous_term_pore}
\langle\nabla \eta\cdot(\nabla\bm u +\nabla\bm u^T)\rangle =\nabla\cdot \eta\left(\nabla \langle \bm u \rangle+(\nabla \langle \bm u \rangle)^T \right)
-\eta\left[\nabla\langle\bm u\rangle^\alpha+(\nabla\langle\bm u\rangle^\alpha)^T\right]\cdot \nabla\epsilon+\frac{1}{V}\int_{A_{\alpha\beta}}(\nabla \hat{\bm u} +\nabla \hat{\bm u}^T )\cdot \bm n_{\alpha\beta}dA.
\end{equation}
The two avearged expressioins are not identical. The integral term in Eq.(\ref{full_viscous_term_pore}) contains the full viscous stress tensor, thereby providing a more complete representation of the viscous resistance exerted by the solid surface on the fluid.
}
\end{remark}

\begin{remark}
By neglecting  the transient, inertial, viscous and   quadratic drag force terms in the averaged momentum equation,   the formulation reduces to the  multiphase Darcy model apart from the inclusion of the interfacial force term
\begin{equation}\label{CH_darcy_simplify}
\frac{\partial \epsilon\langle\phi\rangle^\alpha}{\partial t}+\nabla\cdot\langle\phi\bm u\rangle=\nabla\cdot (M\epsilon\nabla\langle \mu \rangle^\alpha)
\end{equation}
\begin{equation}\label{eq}
\langle\bm u\rangle=-\frac{KK_r}{\langle\eta\rangle^\alpha}\left[\nabla\langle p\rangle^\alpha +\langle\phi \rangle^\alpha\nabla\langle\mu\rangle^\alpha -\langle\rho \rangle^\alpha \bm g \right].
\end{equation}
where $\langle\phi\bm u\rangle$ can be treated as the volume-averaged velocity of the fluid phase, conrresponding to $\langle \bm u_{w}\rangle$ for the wetting phase or $\langle\bm u_{nw}\rangle$  for the  non-wetting phase,as distinguished by the volume fraction $\langle \phi\rangle^\alpha$.
If the viscous dissipation term is included, the volume averaged momentum equation recovers a form of the Brinkman equation
\begin{equation}\label{eq:reduced_brinkman}
\begin{aligned}
\langle\bm u\rangle=&-\frac{KK_r}{\langle\eta\rangle^\alpha}\left[  \nabla\langle p\rangle^\alpha + \langle\phi \rangle^\alpha\nabla\langle\mu\rangle^\alpha -\langle\rho \rangle^\alpha \bm g-\nabla \cdot(\langle\eta\rangle^\alpha ( \nabla \langle\bm u\rangle^\alpha +(\nabla \langle\bm u\rangle^\alpha)^T)  ) \right. \\
&\hspace{1cm}\left. -\frac{1}{\epsilon}\nabla \cdot(\langle\eta\rangle^\alpha ( \langle\bm u\rangle^\alpha\nabla\epsilon +\nabla\epsilon\langle\bm u\rangle^\alpha)  )
 \right].
\end{aligned}
\end{equation}
For sufficiently homogeneous porous media, the last term on the right hand of Eq.(\ref{eq:reduced_brinkman}) can be ignored~\cite{valdes2009validity}, yielding
the classical Brinkman equation.
\end{remark}

\begin{remark}
\textcolor{blue}{
When the porosity $\epsilon$ approaches unity,
  the drag force becomes negligible as  the permeability  becomes infinite. Under these conditions, the  macroscopic equations recovers the pore-scale NSCH equations. Thus, the present model establishes a unified framework that effectively bridges the continuum from the pore scale to the Darcy scale.}
\end{remark}

\begin{remark}
The system free energy functional typically comprises both bulk phase free energy and nonlocal free energy contributions. Rather than employing  a standard double-well potential, we adopt alternative free energy functions that better capture the seepage characteristics of porous media. For instance, by defining the phase chemcial potential as the capillary pressure $P_c=-d\langle f\rangle^\alpha/d\langle\phi\rangle^\alpha$, the bulk free energy functional can be derived inversely from a prescirbed capillary pressure equation.
\end{remark}

\begin{remark}
Carrillo et al~.\cite{carrillo2020multiphase} derived  multiscale governing equations for two-phase flow in porous media from pore-scale hydrodynamic equations. The sum of  their two phase-averaged momentum conservation equations yields a total fluid momentum equation, which is similar to the present averaged momentum equation except in the treatment of the surface tension force and the drag force. Their drag force model includes  only  the linear term in Eq.(\ref{eq:VA_NS_symplify_1}), and the surface tension force is estimated using a continuum surface force formulation. The wetting condition is achieved by modifying  the mean normal vector at the fluid-solid interface. Compared to the Carrrillo's model~.\cite{carrillo2020multiphase}, the averaged Cahn-Hillard-type equation is derived and used to describe the evolution of the saturation $\langle\phi \rangle^\alpha$. The surface tension force is treated in a   potential form  instead of the continuum surface force model~\cite{zhang2021formulations,kim2005continuous}.
\end{remark}

\begin{remark}
\textcolor{blue}{
Compared to existing two-fluid models~\cite{chen2019homogenization},  the averaged equations derived here constitute a  single-fluid model. This approach obviates the need for explicit interfacial bundary conditions, such as expirical capillary pressure relation. Instead,  the interfacal force term appearing in the averaged momentum equation serves as a generalized representation of capillary effects. 
Notably, the wetting boundary condition and specific surface area are intrinsicaly incorporrated into the averaged chemical potential (see Eq.(\ref{eq:VA_Mu}) or (\ref{eq:VA_Mu_close})), offering a novel theoretical pathway to effectively reflect  pore-scale flow characteristics in the macroscopic description.
}

\end{remark}

\subsection{Analysis of the order parameter for a single capillary tube}
\begin{figure}
  \centering
\includegraphics[width=0.45\textwidth]{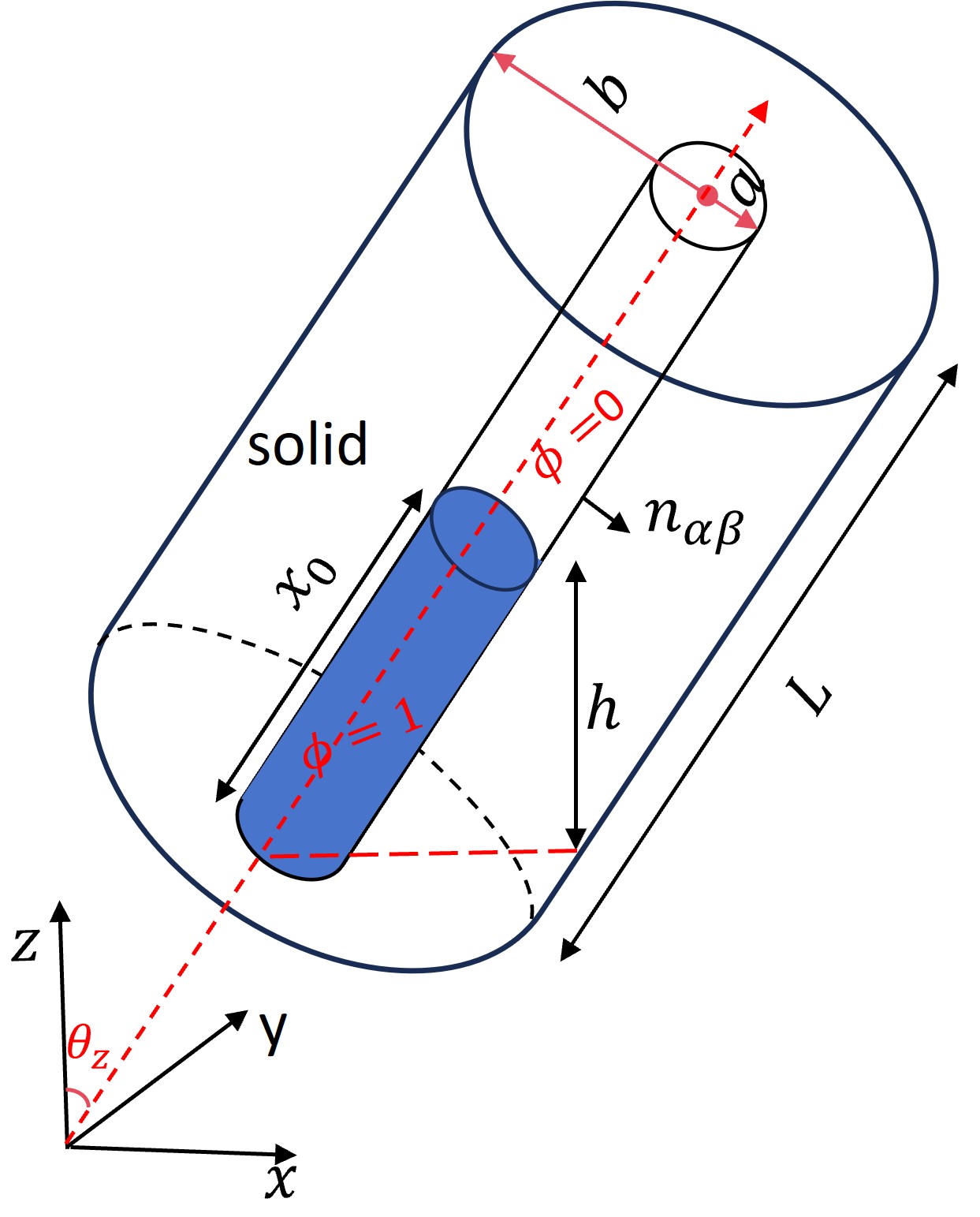}
  \caption{Schematic of a circular capillary tube.}\label{capillarytube}
\end{figure}

\textcolor{red}{
In order to estimate the perturbation order parameter,
we consider a circular capillary tube saturated by wetting fluid and non-wetting fluid, as shown in Fig.(\ref{capillarytube}). The radius of the elementary  tube is $a$ and the radius of the representative volume is $b$. 
The inclination angle of the capillary in the z-axis direction is $\theta_z$.
The  contact angle is assumed to be $\theta$. Based on the Jurin's law, the penetration length is $x_0=2\sigma \cos\theta/((\rho_l-\rho_g)ga\cos\theta_z)$.
As we only consider the homogeneous wettability and $a\ll L$,  the penetration length is not appreciably affected by the axial projection of the curvature radius. }
It is easily built that the porosity is $\epsilon=a^2/b^2$.
Based on the Hagen-Poiseuille law, the relative permeiablity is $\epsilon a^2/8$.

From the definitions of the phase average and intrinsic phase average,  we have $ \langle \phi \rangle^\alpha=x_0/L$.
Recalling $\hat{\phi}=\phi-\langle\phi\rangle^\alpha$, one can obtain
\begin{equation}
\int_{A_{\alpha\beta}}\hat{\phi}\bm n_{\alpha\beta}dA=0,
\end{equation}
\begin{equation}\label{eq:hat_phi_limiting}
\begin{aligned}
\langle(\hat{\phi})^2\rangle^\alpha=&\langle\phi\rangle^\alpha(1-\langle\phi\rangle^\alpha),
\\
 \langle(\hat{\phi})^3\rangle^\alpha=& \langle\phi\rangle^\alpha(1-\langle\phi\rangle^\alpha)(1-2\langle\phi\rangle^\alpha).
\end{aligned}
\end{equation}
  Inserting the above equation to Eq.(\ref{eq:VA_Mu}), the chemical potential becomes
\begin{equation}\label{eq:capillary_mu}
  \langle\mu \rangle^\alpha=
  24\overline{\lambda} (\langle\phi\rangle^\alpha)^3 
  -36\overline{\lambda} (\langle\phi\rangle^\alpha )^2 
  +12\overline{\lambda}  \langle\phi\rangle^\alpha
  -6\epsilon^{-1}\sigma\cos(\theta)a_v \langle\phi\rangle^\alpha(1- \langle\phi\rangle^\alpha)
  -\overline{\kappa} \nabla\cdot\nabla \langle\phi\rangle^\alpha-\frac{\overline{\kappa}}{\epsilon}\nabla\langle\phi \rangle^\alpha \cdot\nabla\epsilon
\end{equation}
\textcolor{red}{where  $\overline{\lambda}$ and $\overline{\kappa}$ denote the  parameters defined at the REV scale. They could differ from the pore-scale counterparts $(\lambda,\kappa)$ and must be scaled accordingly, because dfferent grid resolutions are employed for pore-scale flow and Darcy flow.}
For the fluids saturated the tube at the equilibrium state, the value of the chemical potential should be constant and it is zero for this case.
For the one-dimensional problem,  the profile of $\langle\phi\rangle^\alpha$ can be  written as
\begin{equation}\label{eq}
\langle\phi\rangle^\alpha=\frac{1}{2}-\frac{1}{2}\tanh\left(\frac{2}{\overline{W}}x\right)
\end{equation}
where $\overline{W}=\sqrt{4\overline{\kappa}/3\overline{\lambda}}$.

From Eq.(\ref{eq:capillary_mu}), considering $\epsilon$=const, one may define the system free energy as
\begin{equation}\label{eq}
\langle f\rangle^\alpha=6\overline{\lambda} (\langle \phi \rangle^\alpha)^2 (1-\langle \phi \rangle^\alpha)^2+\frac{\overline{\kappa}}{2}|\nabla \langle \phi \rangle^\alpha  |^2
+6\epsilon^{-1}\sigma\cos(\theta)a_v\left(  \frac{(\langle \phi \rangle^\alpha)^2}{2}-\frac{(\langle \phi \rangle^\alpha)^3}{3}       \right)
\end{equation}
As we consider a plane interface, the last term can be neglected. The  surface tension for  this model  is calculated by
\begin{equation}\label{eq}
 {\sigma}=\int_{-\infty}^{+\infty}\overline{\kappa}  | \nabla \langle \phi\rangle^\alpha |^2dx=\sqrt{\frac{\overline{\lambda}\overline{\kappa}}{3}}.
\end{equation}
\textcolor{red}{It should be noted that  the interface thicnkness is a numerical parameter, typically set to a few grid sizes. At the REV scale, a grid spacing is $\delta r_0$, whereas at the pore scale, it is $\delta l_{\alpha}$. Hence, the numerical grid is inevitably coarsened at the REV scale, while the interfacial tension as a physical property remains consistent across both scales. This scaling leads to the relationships 
\begin{equation}
    \overline{\lambda}/\lambda=6 \frac{r_0}{l_{\alpha}}, \qquad \overline{\kappa}/\kappa=\frac{l_{\alpha}}{r_0}. 
\end{equation}
Using this relation, one can show that  ${\sigma} l/\bar{\kappa}\ll 1$ holds , as required by Eq.(\ref{eq:area_averaged}). }

In addition, based on Eq.(\ref{eq:symplify_hat_phi}), and the restricted condition Eq.(\ref{eq:hat_phi_limiting}),  one can obtain the following relationships
\begin{equation}
\nabla\langle\mu \rangle^\alpha=\bm u_{\phi}\langle\phi\rangle^\alpha, 
\quad
\nabla\langle\phi \rangle^\alpha=\bm s_{\phi} \langle\phi\rangle^\alpha
 \end{equation}
where the vectors $\bm u_\phi$ and $\bm s_\phi$ are still unclosed parameters.

\section{Numerical results}~\label{sec5}
\textcolor{blue}{
To validate the present model and assess its performance,  we develop a lattice Boltzmann method   for the proposed REV-scale governing equations. The  corresponding algorithms are detailed in  Appendix.~\ref{ap:lbm}.
Since the phase-field model contains unclosed parameters, theoretical or experimental validation  of the model is challenging. Therefore, we first neglect the phase field model and interfacial force, and use the present model to simulate a single-phase flow problem with an analytical solution. Then, to demonstrate the performance of the present model, we apply the lattice Boltzmann method to two problems: viscous fingering  and a bubble  rising in porous media, with particular emphasis on illustrating the effects of wettability on interfacial dynamics.}

In simulations, the term $\nabla \cdot(\langle \hat{\phi}\hat{\bm u}\rangle)$ in the averaged CH equation is ignored for simplicity. In Eq.(\ref{eq:VA_Mu_close}), we assume $\bar{\lambda}_3=\bar{\lambda}_2=\lambda$, $\bar{\lambda}_1=\bar{\lambda}_0=0$, $\bm{\bar{\kappa}}=\epsilon\kappa$, $\bar{\bm\sigma }=0$.
The relative permeability proposed by Corey et al.~\cite{brooks1965hydraulic}  is used,
\begin{equation}\label{eq}
  K_{r,w}=S_e^{\frac{2+3\chi}{\chi}}, \quad K_{r,nw}=(1-S_e)^2(1-S_e^{\frac{2+\chi}{\chi}}),  \quad  K_r=K_{r,w}+K_{r,nw}
\end{equation}
where $\chi$ is the pore size distribution index, $S_e=\frac{\langle\phi\rangle^\alpha-S_r}{1-S_r}$ is the  reduced saturation of wetting phase, $S_r$ is the residual saturation of the wetting phase. The surface area per unit volume $a_v$ can be calculated by the Kozeny-Carman equation
\begin{equation}\label{eq}
  K=\frac{\epsilon^3}{c_k(1-\epsilon)^2a_v^2}
\end{equation}
where $c_{k}$ is the Kozeny constant. 
In simulations,  unless other state, $\chi=2$, $S_r=0$, $W=4$, $c_k=2.5$, $\tau_h=0.8$, $\zeta =0.1$ in lattice unit.

\subsection{Poiseuille flow in a homogeneous porous media}
\begin{figure}
     \centering
     \includegraphics[width=0.5\linewidth]{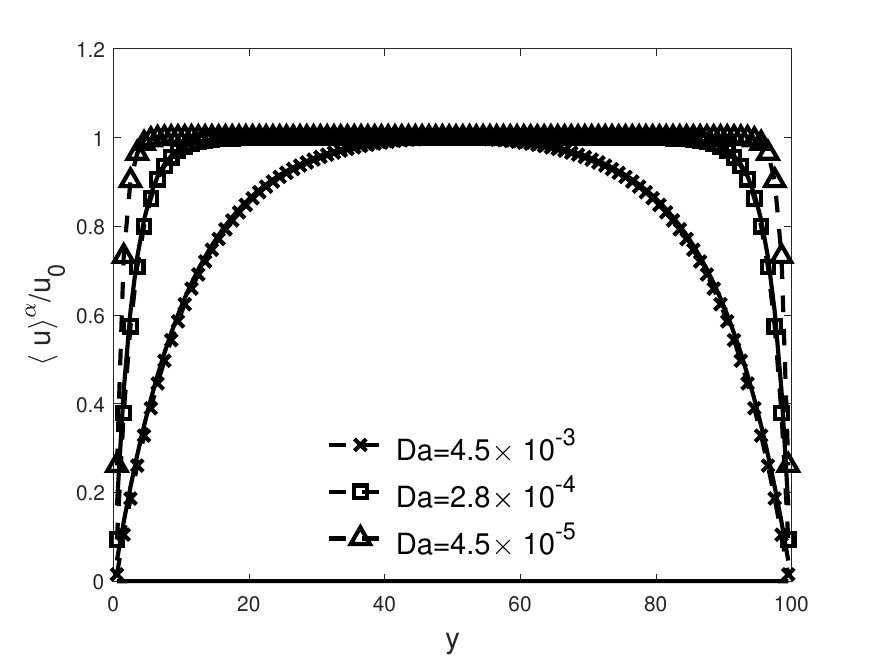}
     \caption{Velocity profiels of the Poiseuille flow for different Darcy numbers. Symbols represent nummerical results and solid lines represent analtyical solutions.}
     \label{fig:bosuyeliu_single}
 \end{figure}

\textcolor{blue}{
For validation purposes, we first examine the hydrodynamic equations by omitting the phase-field equation. Then, the present model reduces to a volume-averaged single-phase Navier-Stokes formulations.
This approach allows direct benchmarking against available analytical solutions before introducing phase field equation.
We consider the Poiseuille flow in a two-dimensional channel of width H filled with a proous medium of porosity $\epsilon$. The flow is driven by a constant force $g$ along the channel direction. 
The nonlinear inertial effect due to the porous medium is ignored. 
As the flow is fully developed along the channel, the streamwise velocity satisfies the following equation
\begin{equation}
\nu_{eff}\frac{\partial^2 \langle u\rangle^\alpha}{\partial y^2}+g-\frac{\epsilon\nu}{K}\langle u\rangle^\alpha=0,
\end{equation}
with $\langle u\rangle^\alpha(x,0)=\langle u\rangle^\alpha(x,H)=0$, and the lateral velocity component is zero everywhere.
The analytical solution can be written as
\begin{equation}
\langle u\rangle^\alpha=\frac{gK}{\epsilon\nu}\left(1- \frac{cosh[r(y-H/2)]}{cosh(rH/2)}\right)
\end{equation}
where $r=\sqrt{\nu\epsilon/K\nu_{eff}}$.
The Reynolds number is defined by $\mathrm{Re}=Hu_0/\nu$, where $u_0$ is the peak velocity of the flow along the centerline.
The simulation is conducted within a computational domain of  $L_x\times L_y=120\times 100$. The porosity is set to be $\epsilon=0.395$. The effective viscosity is equal to the shear viscosity.  The  gravitational acceleration of  $g=1.0\times10^{-5}$ is applied. The permeability is determined by the Darcy number. The Reynolds number is fixed at $\mathrm{Re}=0.29$.
The nonequilibrium extrapolation scheme is applied to the top and bottom walls for non-slip boundary condition.The velocity field is initialized to be zero at each lattice node with a constant density $\rho=1.0$.
Figure~\ref{fig:bosuyeliu_single} compares the simulated velocity profiles with  analytical solutions at different Darcy numbers $\mathrm{Da}$. As shown in Fig.~\ref{fig:bosuyeliu_single}, the effect of the nonlinear drag becomes more significant with higher $\mathrm{Da}$. The numerical results demonstrate good agreement with the corresponding analytical solutions.
}

\subsection{Poiseuille flow in a fluid-porous channel}
\begin{figure}
     \centering
     \includegraphics[width=0.5\linewidth]{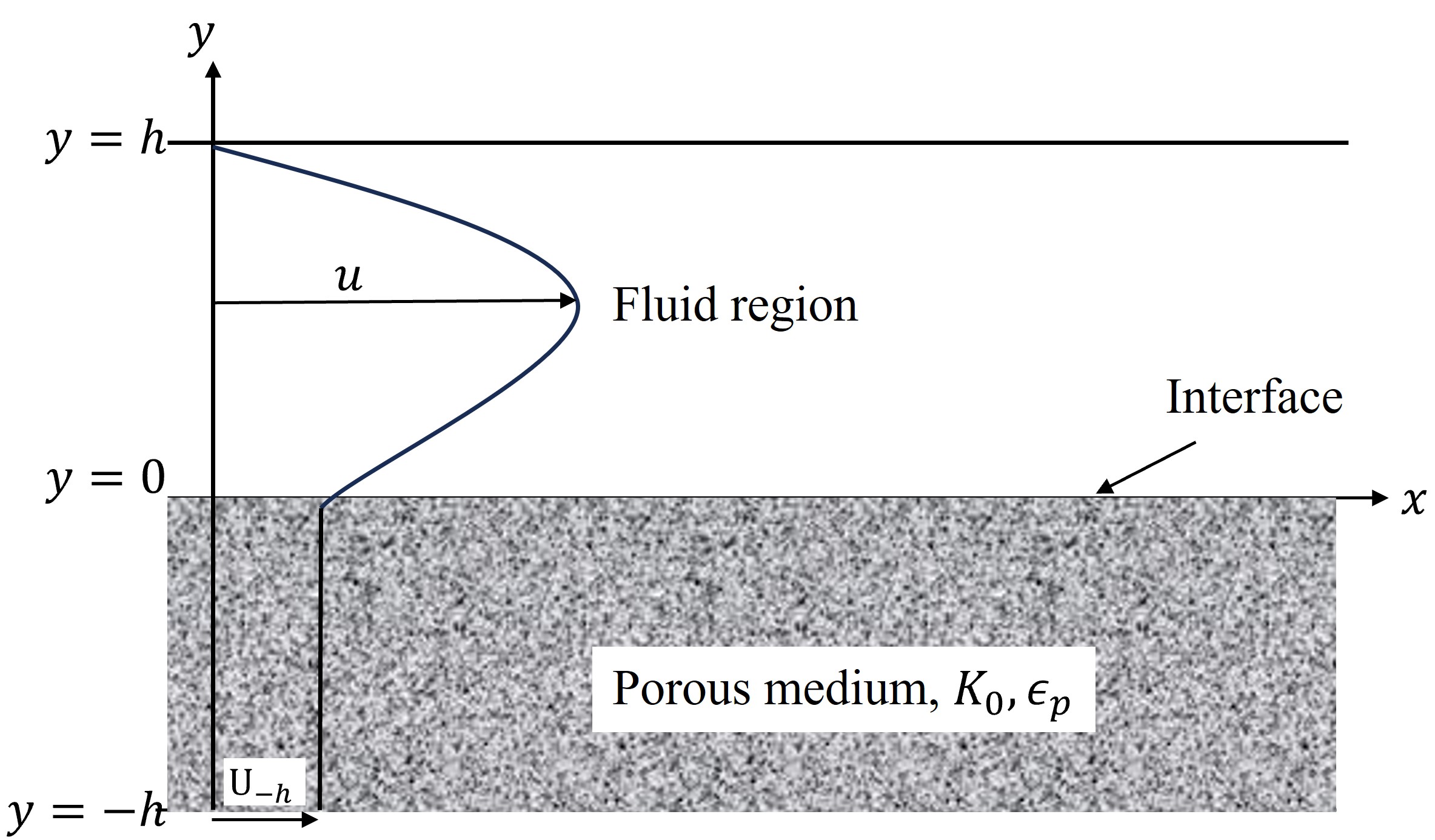}
     \caption{Poiseuille flow in a fluid-porous channel.}
     \label{fig:fluid_porous channel}
 \end{figure}

\textcolor{blue}{
In this test, we consider a Poiseuille flow within a channel partially filled with a porous medium, as shown in Fig.~\ref{fig:fluid_porous channel}. The momentum transport in the free-fluid region is governed by the Stokes equations, while flow in the porous medium is described by the Darcy-Brinkman equation. The porosity and permeability of the porous medium remote from the interface are denoted by $\epsilon_p$ and $K_0$, respectively. 
A  no-slip condition is applied at  the top wall.
In the porous medium distant from the interface, the 
Darcy velocity $U_{-h}=-K_0/\eta\nabla \langle p\rangle^\alpha $ is enforced.
At the interface, the velocity should be continuous while the stress boundary condition displays a discontinuity described by 
\begin{equation}
\left.\eta \frac{\partial u}{\partial y}\right|_{y=0}-\left.\eta_{eff}\frac{\partial U}{\partial y}\right|_{y=0}=-\frac{\beta_p}{\sqrt{K_0}}u(0),
\end{equation}
where $\beta_p$ represents the stress jump coefficient, a value determined experimentally. Goyeau et al. suggested that $\beta$ is intrinsically linked to the continous spatial variations of the porous structure and velocity within the transition zone.
Considering   the continuous porosity transition from  $\epsilon_p$ to unity near the interface,  the porosity field is modeled as 
\begin{equation}
\epsilon(x,y)=\frac{\epsilon_p+1}{2}+\frac{1-\epsilon_p}{2}\tanh\left(\frac{y}{W} \right),
\end{equation}
where $W$ parameterizes the length scale of this transition.
The thickness $W$ decreases when the Darcy number decreases. 
For this problem,  an  analytical solution proposed by Chandesris~\cite{chandesris2006boundary}  is given by 
\begin{equation}
\langle u\rangle =U_{-h}\left(1+\frac{\zeta^2-2}{2(1+\zeta\sqrt{\epsilon_p})}\exp\left(\sqrt{\frac{\epsilon_p}{K_0}}y \right) \right), \qquad  y\leq 0
\end{equation}
\begin{equation}
\langle u\rangle =U_{-h}\left(
\frac{\zeta(\zeta+2\sqrt{\epsilon_p})}{2(1+\zeta\sqrt{\epsilon_p})}
+
\sqrt{\epsilon_p}\frac{\zeta^2-2}{2(1+\zeta\sqrt{\epsilon_p})}
\left(\frac{y}{\sqrt{K_0}} \right)
-
\frac{1}{2}\left(\frac{y}{\sqrt{K_0}} \right)^2
\right), \qquad  0\leq y\leq h,
\end{equation}
where $\zeta=1/\sqrt{\mathrm{Da}}$, $\mathrm{Da}=K_0/h^2$ is the Darcy number.}

\begin{figure}[htbp]
  \centering
 
\subfloat[$\epsilon_p=0.2$]{
\includegraphics[width=0.33\textwidth]{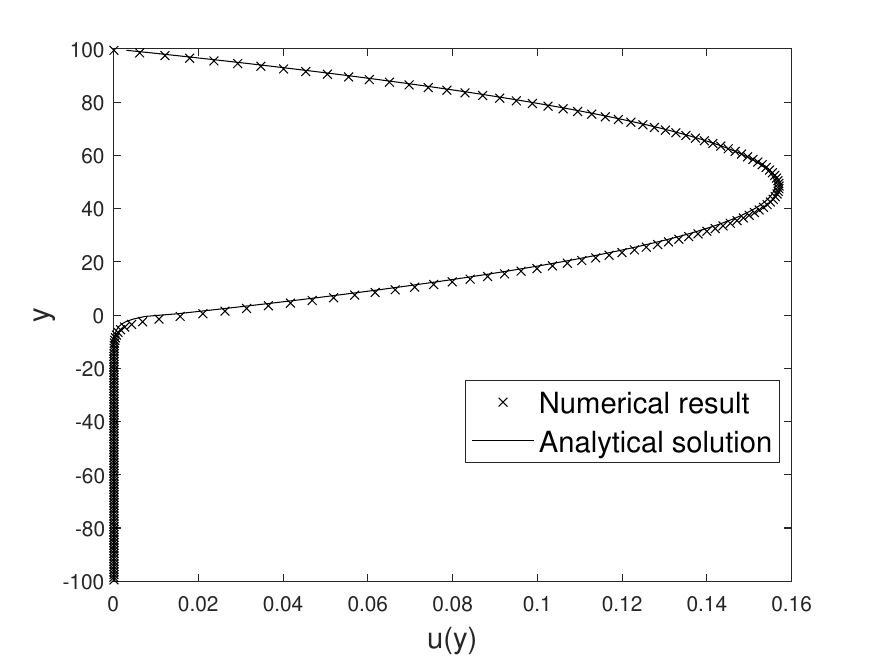}}
\hfill
\subfloat[$\epsilon_p=0.4$]{%
\includegraphics[width=0.33\textwidth]{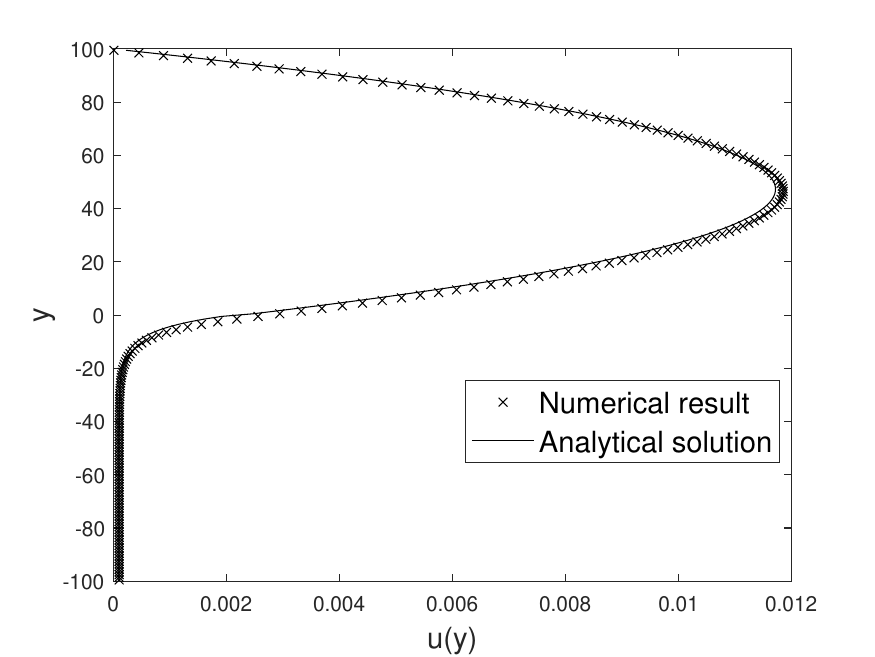}}
\hfill
\subfloat[$\epsilon_p=0.58$]{%
\includegraphics[width=0.33\textwidth]{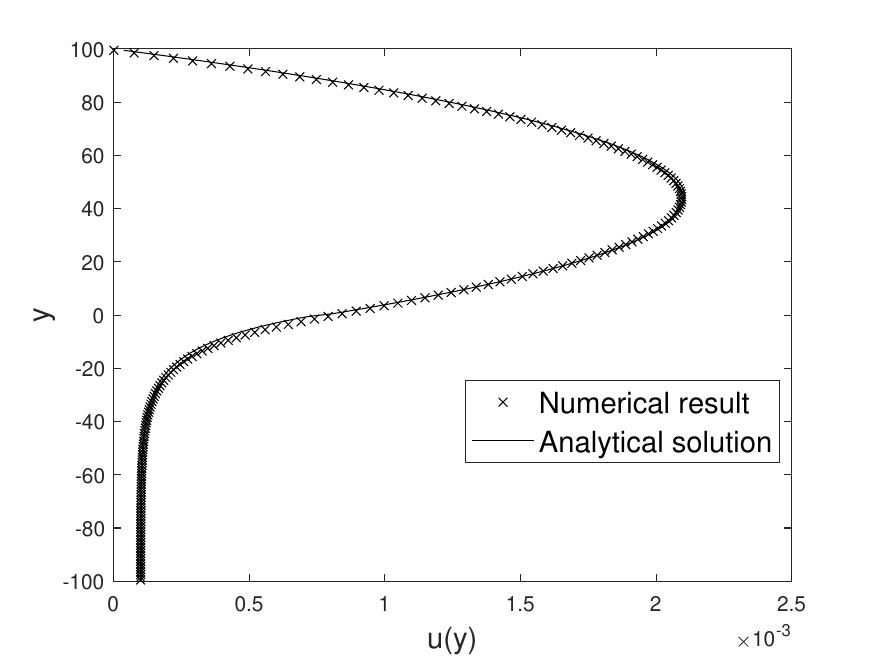}}

\caption{Velocity profiles for  Poiseuille flow in a fluid-porous channel at different porosities.} 
\label{fig:comparison_fluid_porous channel}
\end{figure}

\textcolor{blue}{
In simulations, the computational domain is discretized using a uniform mesh of $L_x\times L_y=100\times 200$. The height of the free-fluid region is $h=100$. The porosity in the porous medium region is set to be $\epsilon_p=0.4$. The other parameters are set as $\nu_{eff}=\nu=0.1$, $U_{-h}=0.0001$. 
For boundary conditions, the nonequilibrium extrapolation scheme is applied to the top wall to implement the non-slip condition, and the mirror reflection boundary condition is applied to the bottom wall. 
Flow is driven by a body force applied in the streamwise direction, which is calibrated such that the Darcy velocity in the region far from the interface satisfy $U_{-h}$. Following the treatment~\cite{goyeau2003momentum},  $W$ is adjusted to match the interfacial velocity to the analytical solution and is set to $4$.
Figure~\ref{fig:comparison_fluid_porous channel} presents the velocity profiles. There is a good agreement between the numerical results and the analytical solution.  This confirms the capability of the model to simulate flows in heterogeneous porous media.}

\subsection{\textcolor{blue}{Buckley-Leverett problem}}
\textcolor{blue}{
The Buckley-Leverett theory has been widely used in the literature to analyze the structure of two-phase flows through both homogeneous and heterogeneous media. It is also used as a reference to validate numerical method and to evaluate their accuracy in capturing shocks. Here, we consider  a one-dimenional domain aligned with the $x$-direction. The wetting phase is injected from the left into an uniform domain saturated with nonwetting phase. In the right side of the domain, a fixed pressure condition is imposed.   
Both capillary and gravity effects are neglected. 
Based on the standard two-phase Darcy equations, the classical Buckley-Leverett equation is expressed as 
\begin{equation}
\begin{aligned}
    \epsilon\frac{\partial S_w}{\partial t}+u_t\frac{\partial \mathrm{f_w}(S_w)}{\partial x}=0, \\
    S_w(x=0,t)=1-S_{r,nw},\qquad S_w(x=L_x,t)=S_{wi},
\end{aligned}
\end{equation}
where $S_w$ denotes the wetting phase saturation, $S_{wi}$ and $S_{r,nw}$ denotes the initial wetting phase saturation and residual non-wetting phase saturation, respectively; $u_t$ is Darcy velocity . The fractional flow function $\mathrm{f_w}$ is defined by 
\begin{equation}
    \mathrm{f_w}=\frac{k_{rw}/\eta_w}{k_{rw}/\eta_w+k_{rnw}/\eta_{nw}},
\end{equation}
}

\textcolor{blue}{
The analytical solution for the saturation is given by
\begin{equation}
 S_w(x,t)=
 \begin{cases}
 S_{wf} & \text{for} \quad
 x=x_f(t), \\
 \left(\frac{u_tt}{\epsilon}\mathrm{f_w}'(S_{wf})\right)^{-1} & \text{for} \quad x<x_f(t), \\
 S_{wi} & \text{for} \quad x>x_f(t) ,
 \end{cases}
\end{equation}
where $x_f(t)=u_t t \mathrm{f_w}'(S_{wf})/\epsilon $ is the leading edge of the wetting phase front, which is calculated by the Welge tangent method.
To close the averaged phase-field equaiton and ensure alignment with the Buckley-Leverett theory, the phase velocity term $\langle\phi\bm u\rangle$ in Eq.(\ref{CH_darcy_simplify}) is reformulated as $\epsilon\langle \phi\rangle^\alpha\bm \langle u\rangle^\alpha_{eff}$. Here, the effective velocity is defined as $\langle u\rangle^\alpha_{eff}=\langle\bm u\rangle^\alpha \mathrm{f_w}/\langle \phi\rangle^\alpha$. Furthermore, it should be noted that $S_w$ and $\langle\phi\rangle^\alpha$ are equivalent, Here, we retain both symbols to ensure the clarity of the derivation and to distinguish our phase-field formulation from the conventional Darcy’s equation.
In simulations, the computation domain is discretized into a grid of size $L_y\times L_x=2\times 600$, the injection velocity is $0.0001$, and the outlet pressure is fixed at zero. The other parameters are set as $S_{wi}=0$, $S_{r,nw}=0$, $\epsilon=0.5$, $\chi=2$, $\rho_1=\rho_{2}=1$, $\sigma=1\times 10^{-5}$, $M_{eff}=0.01$, $\eta_w=0.05$, $\theta=90^\circ$.}

\begin{figure}[htbp]
  \centering
\subfloat[$\eta_{nw}/\eta_{w}=5$]{
\includegraphics[width=0.33\textwidth]{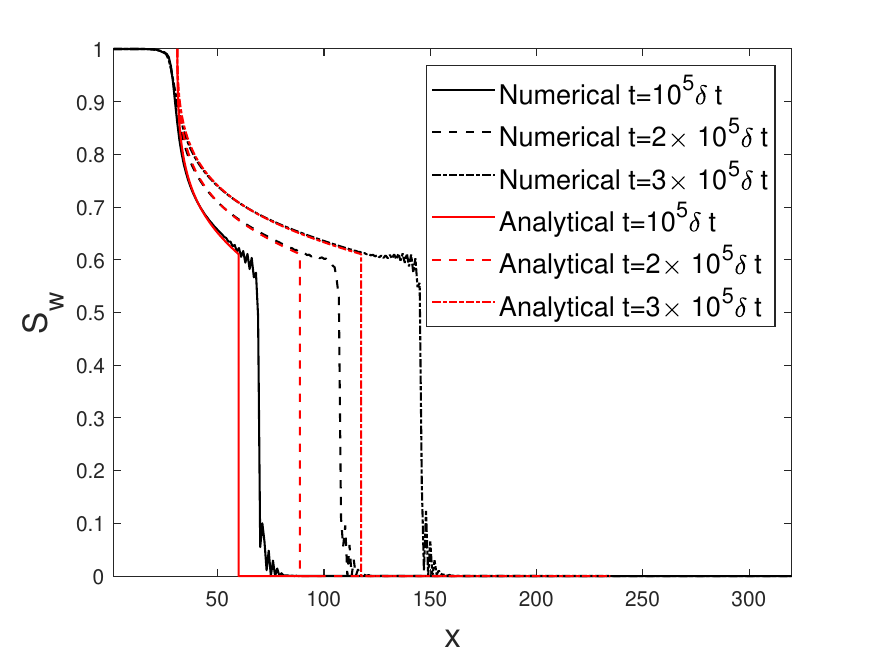}}
\hfill
\subfloat[$\eta_{nw}/\eta_{w}=10$]{%
\includegraphics[width=0.33\textwidth]{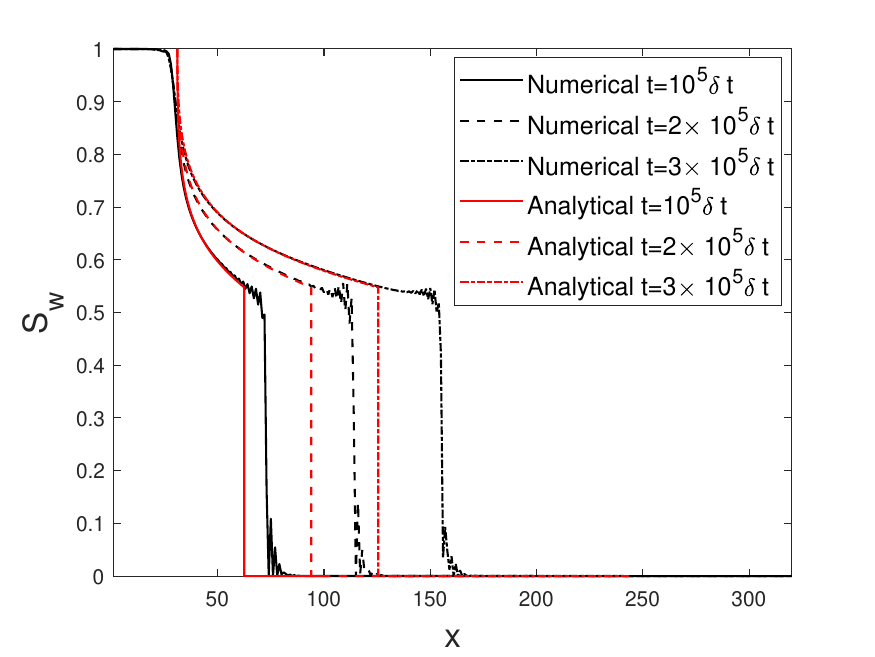}}
\hfill
\subfloat[$\eta_{nw}/\eta_{w}=20$]{%
\includegraphics[width=0.33\textwidth]{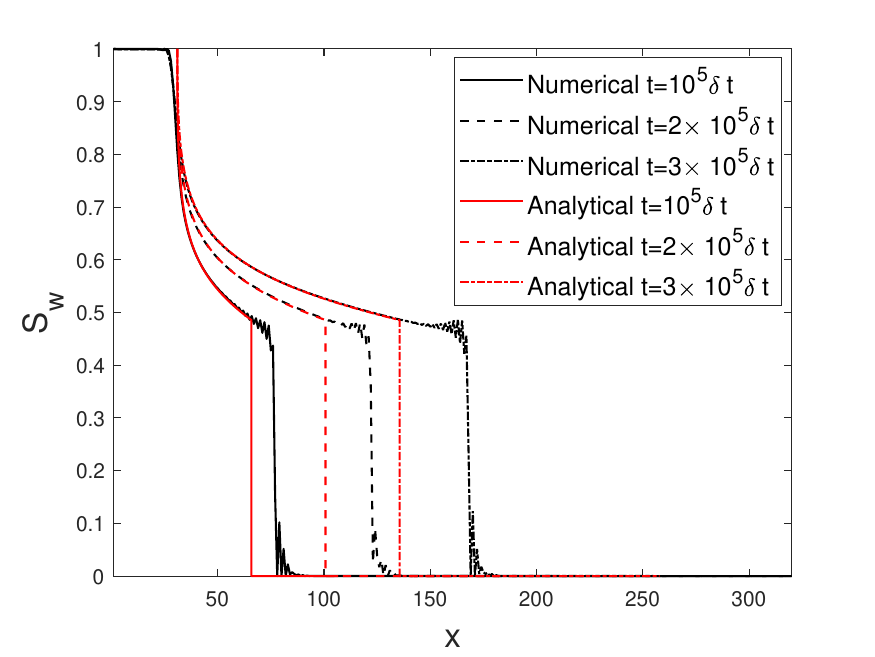}}

\caption{Saturation profiles along the x-direction for the Buckley-Leverett problem.} 
\label{fig:comparison_vis_BL}
\end{figure}

\textcolor{blue}{
Figure~\ref{fig:comparison_vis_BL} presents a comparison between the simulated saturation profiles and the corresponding analytical solutions.
 The shock-front saturation is observed to decreases as the viscosity ratio increases. Based on the Welge's construction, the theoretical front saturations are  $0.611$, $0.583$ and $0.486$ for viscosity ratios of $5$, $10$ and $20$, respectively. In comparison, the corresponding simulated front saturations are   $0.60$, $0.55$ and $0.48$. The maximum relative error  for the front saturation is $5.6\%$.
The results exhibit slight numerical  oscillations in the vicinity of  the shock front. This phenomenon  primarily stems from the inherent nature of the Lattice Boltzmann  scheme, which shares similarities with central-difference-based methods lacking robust flux limiters. Despite these artifacts, the numerical results agree well with the analytical solutions  in the upstream region of the shock wave front.}

\textcolor{blue}{
It is also observed that the simulated shock front  advances faster than the analytical solution, which may be likely attributable to numerical diffusion and the specific choice of the phase-field mobility coefficient. Specifically, localized saturation fluctuations  generate spurious convective velocities, which accelerate the shock front beyond its theoretical propagation speed (where the velocity should ideally remain zero behind the front). Although adjusting the mobility can improve the match, the use of a Single-Relaxation-Time collision operator in the current study imposes stability constraints. Overall, the numerical results demonstrate the model's capability to capture the fundamental dynamics of  two-phase displacement.}

\subsection{Viscous fingering in  porous media}
\begin{figure}[htbp]
  \centering
\subfloat[ $\theta=30^\circ$]{\includegraphics[width=0.8\textwidth,trim=0 0 0 0, clip]{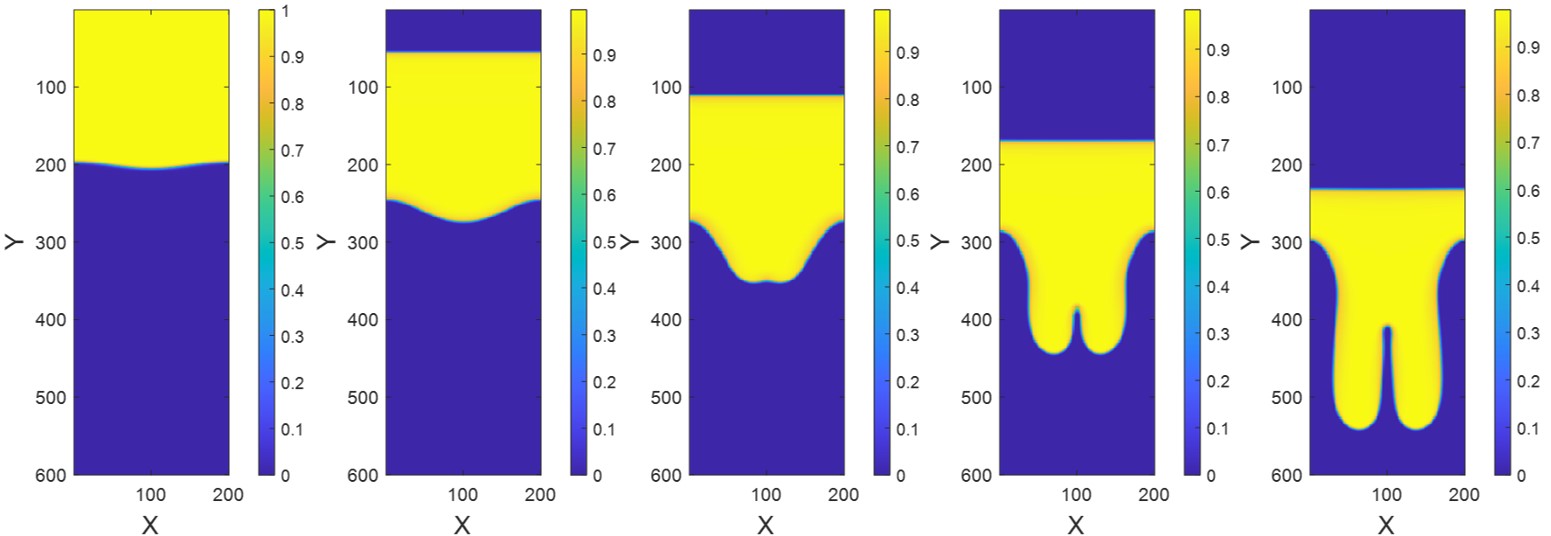}}\\
\subfloat[$\theta=90^\circ$]{\includegraphics[width=0.8\textwidth,trim=0 0 0 0, clip]{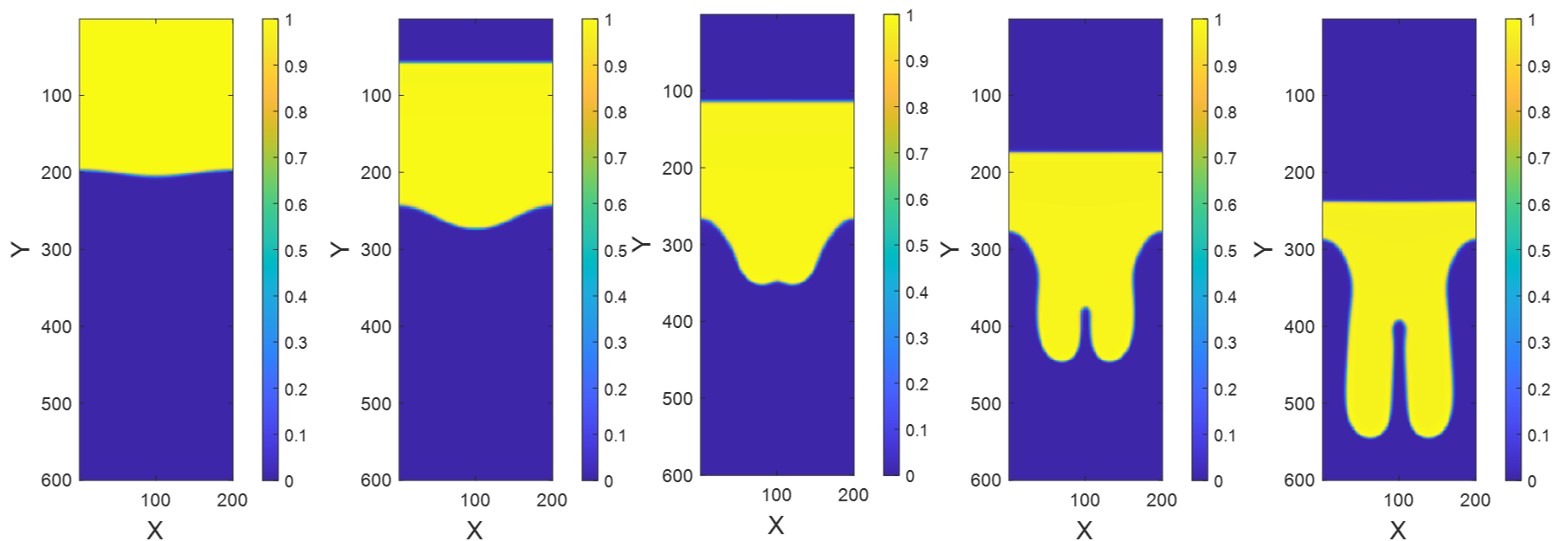}}\\
\subfloat[ $\theta=120^\circ$]{\includegraphics[width=0.8\textwidth,trim=0 0 0 0, clip]{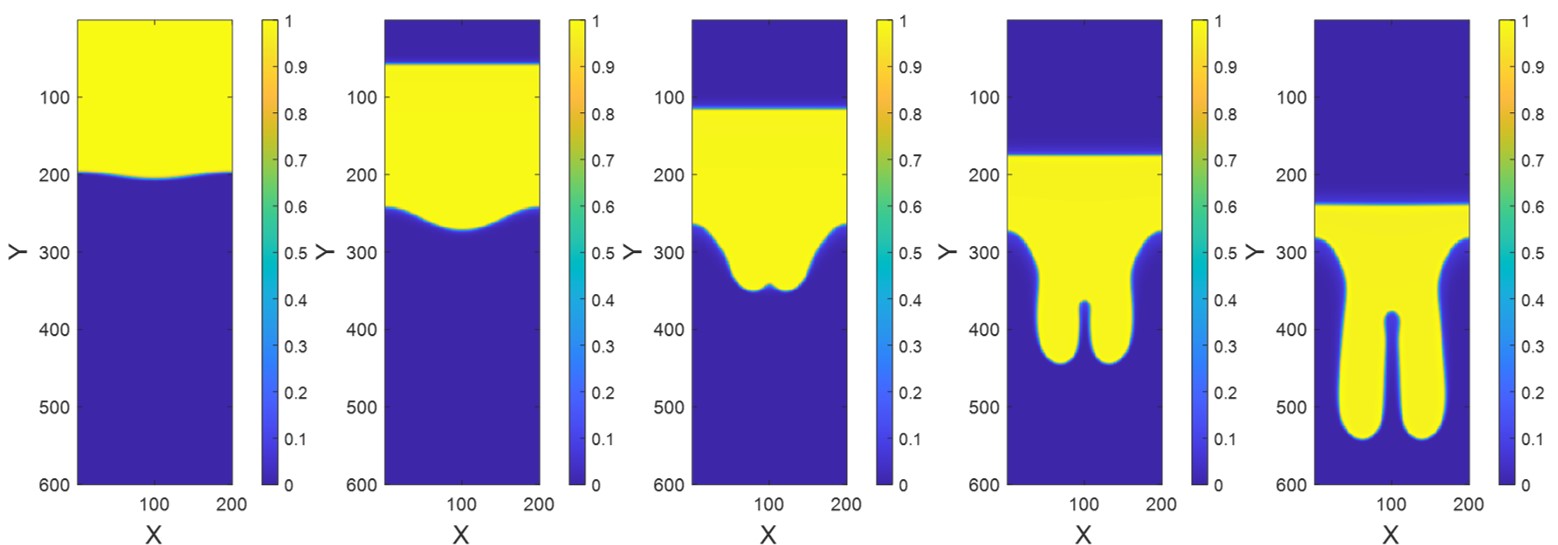}}\\
  \caption{ Viscous fingering with different contact angles and $\eta_2/\eta_1=16.7$. From left to right, $t\sqrt{g}/\sqrt{L_x}=0, 100, 200,300, 400.$ }\label{VF_da-4_vis16}
\end{figure}

\begin{figure}[htbp]
  \centering
\subfloat[ $\theta=30^\circ$]{\includegraphics[width=0.8\textwidth,trim=0 0 0 0, clip]{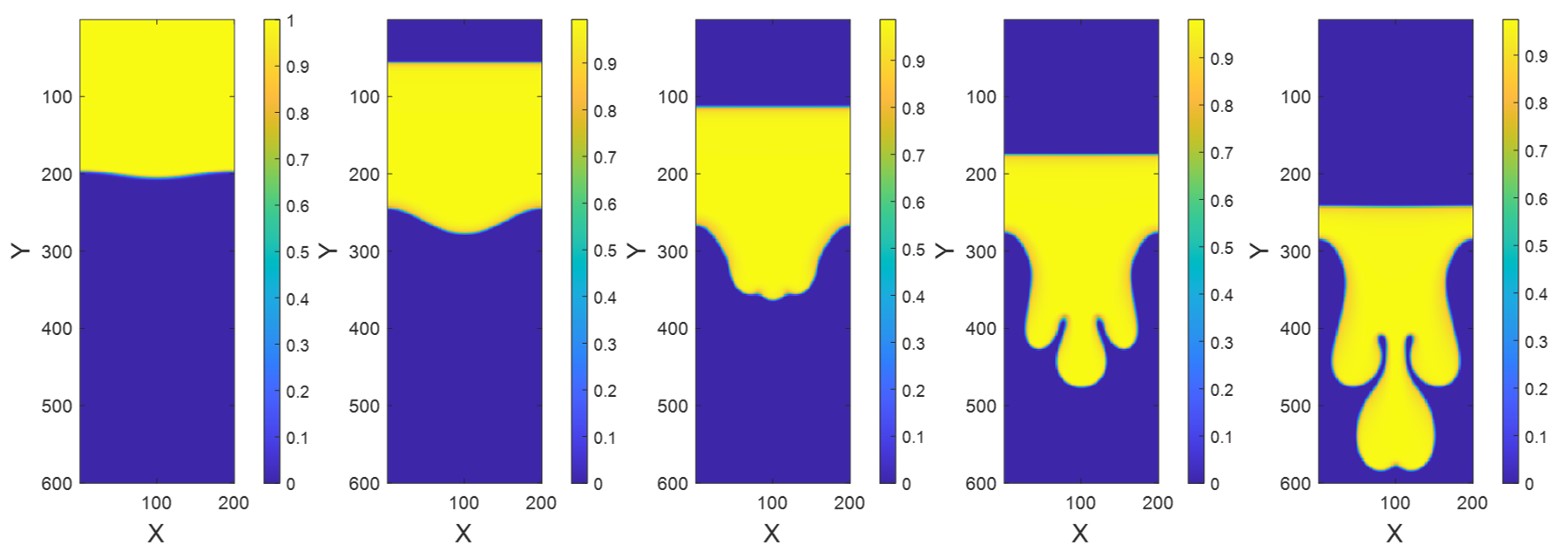}}\\
\subfloat[$\theta=90^\circ$]{\includegraphics[width=0.8\textwidth,trim=0 0 0 0, clip]{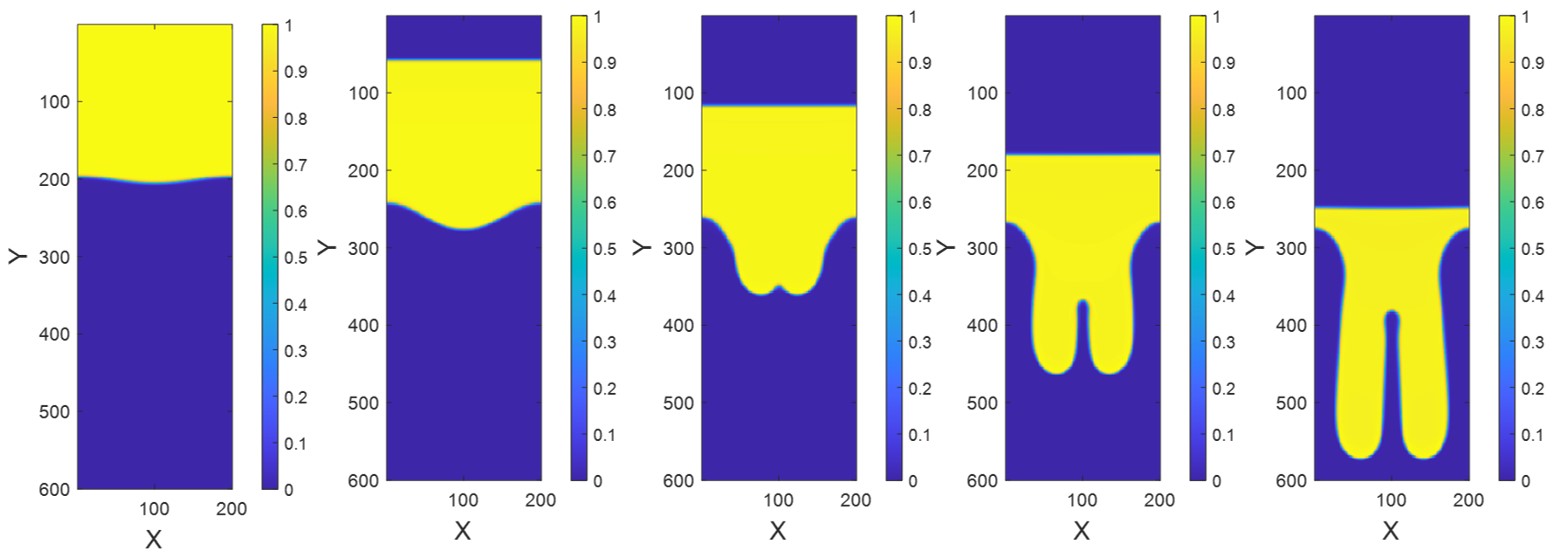}}\\
\subfloat[ $\theta=120^\circ$]{\includegraphics[width=0.8\textwidth,trim=0 0 0 0, clip]{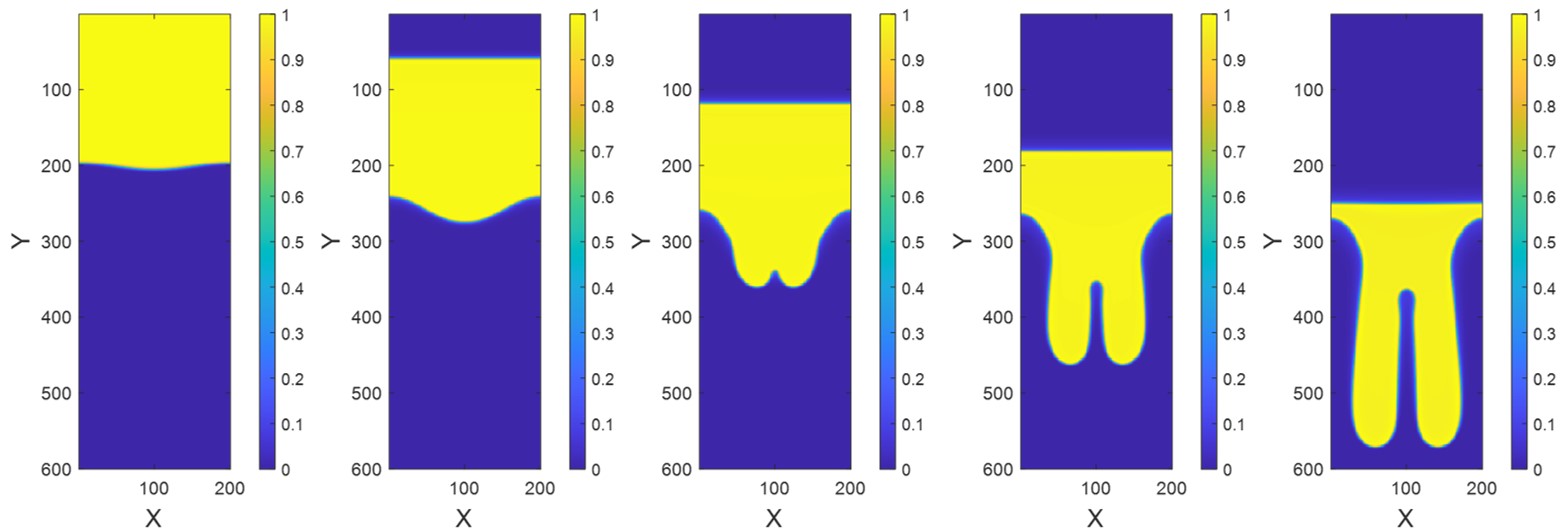}}\\
  \caption{Viscous fingering with different contact angles and $\eta_2/\eta_1=41.7$.  From left to right, $t\sqrt{g/L_x}=0, 100, 200,300, 400$.}\label{VF_da-4_vis40}
\end{figure}

Finally,  we conduct a qualitative simulation of  viscous fingering behavior in porous media, focusing on the effect of wettabiity on the instability. 
Many works on such viscous fingering have been reported~\cite{homsy1987viscous, hosseinzadehsadati2022impact,sorbie2020modelling}.
An important quantity governing viscous instability is  the phase mobility ratio, $M=\lambda_1/\lambda_{2}=(k_{r1}\eta_2)/(k_{r2}\eta_1)$.
Now, we consider that both fluids have the same density and different viscosity ratios $\eta_1\ll \eta_2$. 
\textcolor{blue}{It is worth noting that the adoption of a relatively high viscosity ratio suggests a potential contribution from the term $\hat{\bm R}_{\eta}$. However, as indicated in Eq.(\ref{eq:NSCH_homogenous}), the viscous terms become negligible  when the Darcy number is sufficiently small. Therefore, the influence of $\hat{\bm R}_{\eta}$ can be disregarded for this specific case.}
The porous media is homogeneous and the permeability $K$ is a constant in space.
Periodic boundary conditions are applied to all boundaries. The flow driven by the external force $g$.
The initial velocity is set to zero in the whole domain. The computational domain is discretized by a uniform mesh $L_x\times L_y=200\times 600$.
the porosity is set to be $\epsilon=0.3$, Darcy number is set as $Da=K/L_x^2=6.25\times 10^{-6}$.
 The other parameters are set as $\rho_1=\rho_2=1$, $\eta_2=0.167$,  $\sigma=0.001$, $ g=3.2\times10^{-5}$, $W=4$.  The viscosity of the displacing phase is determined by the viscosity ratio.
 To  form viscous fingering quickly,  the  initial interface between the two fluids is given by
 \begin{equation}\label{eq}
 \langle\phi \rangle^\alpha=\frac{1}{2}-\frac{1}{2}\tanh\left(\frac{2(y-( L_x-4\cos(2\pi x/L_x)))}{W}\right).
 \end{equation}
Figure~\ref{VF_da-4_vis16} shows the snapshots of $\langle\phi\rangle^\alpha$  at different contact angles and  the viscosity ratio $\eta_2/\eta_1=16.7$.
It can be seen that these fingers  patterns are almost similar. The fingers develop into a mushroom shape as they grow, and split to form second-generation fingers when the width of the growing interface is big enough.
We further simulated this case but a large viscosity ratio  $\eta_2/\eta_1=41.7$ is used,  as shown in Fig.~\ref{VF_da-4_vis40}.
When $\theta=30^\circ$, the second and third generations of splitting occur. When  $\theta=90^\circ$ and $120^\circ$, only second generation of splitting is observed.
This implies that the hydrophilic property of porous media strongly disperses the incipient viscous fingers and effectively hinders the fingers from forming.
These results are qualitatively consistent with   the experimental observations in Beteta et al.~\cite{beteta2024immiscible}.

\section{Conclusions}~\label{sec6}
\textcolor{blue}{This paper presents a rigorous macroscopic framewrok for two-phase incompressible flows in porous media, derived by upscaling the pore-scale Navier–Stokes–Cahn–Hilliard equations using the volume averaging technique. The pivotal novelty of this work lies in the establishment of a unified single-fluid model that seamlessly bridges pore- and Darcy-scale representations. Specifically, the proposed model consistently recovers the multiphase flow physics at the pore scale while reducing to a generalized Darcy-scale formulation with thermodynamically consistent constitutive relations.}

\textcolor{blue}{
A key feature of the model is the volume-averaged chemical potential, which provides a formal derivation of capillary forces within the macroscopic momentum equation. This approach yields a generalized formulation for capillary pressure that intrinsically couples wettability through the chemical potential. Consequently, it offers a new theoretical pathway to incorporate wettability effects into a unified, scale-bridging formalism, enhancing the representation of interfacial dynamics in porous systems.}

\textcolor{blue}{
To implement the derived governing equations, a lattice Boltzmann method was developed. Initially, the phase field equation is neglected to examine the accuracy of the hydrodynamic equations.
Poiseuille flow in a homogeneous porous medium and flow in a fluid-porous coupled channel are simulated, which show excellent agreewell with the theretical solutions. Subsequently,incorporating the phase field equation, we simulate the Buckley-Leverett problem to validate the model's capability to capture the transitional mixing region between phases. Finally, we investigate viscous fingering in porous media, elucidating the influence of wettability on interfacial dynamics.}

\textcolor{blue}{
It is woth noting that centain closure parameters, which may be critical in realistic flows,  have been simplified in this study and should be further identified through experimental data 
or theoretical refinement. While a capillary tube approximation was adopted here to simplify the chemical potential, actual porous structures are significantly more complex. Therefore, a more rigorous evaluation of the closure coefficients ($\bm a_{\phi}$, $\bm b_{\phi}$, $c_\phi$ )  for general geometries remains essential. Future investigations should also focus on establishing explicit relationships between the present interfacial force and conventional capillary pressure models, such as the Leverett J-function. Additionally, the influence of different interpolation schemes for mixed-fluid viscosity and relative permeability on the interfacial behavior merits further systematic analysis. In summary, this work provides a foundational, scale-bridging framework that establishes a promising theoretical basis for the study of complex multiphase transport in porous media.  }

\begin{appendix}
\section{The deviation equation for the Navier-Stokes equations with matched density and viscosity ratios}
In order to close the volume average Navier-Stokes equations, the equations for the spatial deviation $\hat{\bm u}$ are derived.
For brevity, we consider the  fluids with matched density and viscosity ratios.
By subtracting  Eq.(\ref{eq:vA_NS_final}) from Eq.(\ref{eq:NS}), we can obtain the equation governing  the deviation momentum equation
\begin{equation}\label{eq:ap_momentum_hatu}
 \begin{aligned}
 &\partial_t(  \langle\rho \rangle^\alpha  \hat{\bm u}   )
 +\nabla\cdot\left[  \langle\rho \rangle^\alpha  \langle\bm u \rangle^\alpha\hat{\bm u} + \langle\rho \rangle^\alpha \hat{\bm u} \langle\bm u \rangle^\alpha-  \langle\rho\rangle^\alpha\langle\hat{\bm u}\hat{\bm u}\rangle^\alpha
  \right] -\epsilon^{-1} \left[   \langle\rho \rangle^\alpha  \langle\bm u \rangle^\alpha\langle\bm u \rangle^\alpha + \langle\rho \rangle^\alpha  \langle\hat{\bm u}\hat{\bm u} \rangle^\alpha    \right]\cdot \nabla\epsilon
  \\
&=  -\nabla\hat{p}+\nabla\cdot(\langle \eta \rangle^\alpha (\nabla\hat{\bm u}+\nabla\hat{\bm u}^T ))
 -\epsilon^{-1}\nabla\cdot(\langle \eta \rangle^\alpha (\langle{\bm u}\rangle^\alpha\nabla\epsilon+\nabla\epsilon\langle\bm u\rangle^\alpha )) \\
 &
 -\langle\phi \rangle^\alpha\nabla\hat{\mu}-\hat{\phi}\nabla \langle \mu \rangle^\alpha -\hat{\phi}\nabla\hat{\mu}+\hat{\rho}\bm g
 \\
 &
-\frac{\epsilon^{-1}}{ V}\int_{A_{\alpha\beta}}\left(-\hat{p}\bm I+\eta(\nabla\hat{\bm u}+\nabla\hat{\bm u}^T)  \right) \cdot\bm n_{\alpha\beta}dA
\end{aligned}
\end{equation}
Recall the assumption of length-scale separation, we can realize some simplifications. The  convective terms on the left hand side (LHS) of Eq.(\ref{eq:ap_momentum_hatu}) can be estimated by
\begin{equation}\label{eq}
\begin{aligned}
\hat{\bm u}\cdot\nabla\langle\rho\rangle^\alpha\langle\bm u\rangle^\alpha=O\left( \frac{\langle\rho\rangle^\alpha(\langle\bm u\rangle^\alpha)^2}{L}\right), \\
\hat{\bm u}\nabla\cdot\langle\rho\rangle^\alpha\langle\bm u\rangle^\alpha=O\left( \frac{\langle\rho\rangle^\alpha(\langle\bm u\rangle^\alpha)^2}{L}\right), \\
\nabla\cdot\langle\rho\rangle^\alpha\langle\hat{\bm u}\hat{\bm u}\rangle^\alpha=O\left( \frac{\langle\rho\rangle^\alpha(\langle\bm u\rangle^\alpha)^2}{L}\right), \\
 \epsilon^{-1}\langle\rho\rangle^\alpha\langle\hat{\bm u}\hat{\bm u}\rangle^\alpha\cdot\nabla\epsilon=O\left( \frac{\langle\rho\rangle^\alpha(\langle\bm u\rangle^\alpha)^2}{L}\right),\\
 \langle\rho\rangle^\alpha\langle\bm u\rangle^\alpha\cdot\nabla\hat{\bm u}=O\left( \frac{\langle\rho\rangle^\alpha(\langle\bm u\rangle^\alpha)^2}{l}\right).
\end{aligned}
\end{equation}
Therefore, from the length-scale constraint $l\ll L$, we obtain
\begin{equation}\label{eq}
\text{LHS}=\partial_t( \langle\rho \rangle^\alpha \hat{\bm u})+ \langle\rho\rangle^\alpha\langle\bm u\rangle^\alpha\cdot\nabla\hat{\bm u}.
\end{equation}
Similarly,  the terms in the right hand side (RHS) of Eq.(\ref{eq:ap_momentum_hatu}) can be estimated by
\begin{equation}\label{eq}
  \begin{aligned}
  \nabla\cdot(\langle \eta \rangle^\alpha (\nabla\hat{\bm u}+\nabla\hat{\bm u}^T )) &=O\left(\frac{\langle \eta \rangle^\alpha \langle \bm u\rangle^\alpha }{l^2}\right),  \\
    \epsilon^{-1}\nabla\cdot(\langle \eta \rangle^\alpha (\nabla\epsilon\langle\bm u\rangle^\alpha+\langle\bm u\rangle^\alpha\nabla\epsilon ))&=O\left(\frac{\langle \eta \rangle^\alpha \langle \bm u\rangle^\alpha }{L^2}\right),\\
    \frac{\epsilon^{-1}}{ V}\int_{A_{\alpha\beta}}\left( \eta(\nabla\hat{\bm u}+\nabla\hat{\bm u}^T)  \right) \cdot\bm n_{\alpha\beta}dA  &=O\left(\frac{\langle \eta \rangle^\alpha \langle \bm u\rangle^\alpha }{l^2}\right).
  \end{aligned}
\end{equation}
With the local equilibrium hypothesis, we deduce that
\begin{equation}\label{eq}
  \begin{aligned}
\text{RHS}=&  -\nabla\hat{p}+\nabla\cdot(\langle \eta \rangle^\alpha (\nabla\hat{\bm u}+\nabla\hat{\bm u}^T ))
-\hat{\phi}\nabla \langle \mu \rangle^\alpha +\hat{\rho}\bm g \\
& -\frac{\epsilon^{-1}}{ V}\int_{A_{\alpha\beta}}\left(-\hat{p}\bm I+\eta(\nabla\hat{\bm u}+\nabla\hat{\bm u}^T)  \right) \cdot\bm n_{\alpha\beta}dA
\end{aligned}
\end{equation}
Finally, with these simplifications, the closure equation for the momentum balance takes the form
\begin{equation}\label{eq:ap:simplifcaiton}
  \begin{aligned}
  \partial_t( \langle\rho \rangle^\alpha \hat{\bm u})+ \langle\rho\rangle^\alpha\langle\bm u\rangle^\alpha\cdot\nabla\hat{\bm u}=& -\nabla\hat{p}+\nabla\cdot(\langle \eta \rangle^\alpha (\nabla\hat{\bm u}+\nabla\hat{\bm u}^T ))
-\hat{\phi}\nabla \langle \mu \rangle^\alpha  +\hat{\rho}\bm g
 \\
 &
-\frac{\epsilon^{-1}}{ V}\int_{A_{\alpha\beta}}\left(-\hat{p}\bm I+\eta(\nabla\hat{\bm u}+\nabla\hat{\bm u}^T)  \right) \cdot\bm n_{\alpha\beta}dA
  \end{aligned}
\end{equation}
As we focus on the evaluation of the order of magnitude of the terms, we can found
\begin{equation}\label{eq}
\begin{aligned}
  \partial_t( \langle\rho \rangle^\alpha \hat{\bm u})/\nabla\cdot(\langle \eta \rangle^\alpha (\nabla\hat{\bm u}+\nabla\hat{\bm u}^T ))=O\left(\frac{  \langle\rho \rangle^\alpha  \langle\bm u \rangle^\alpha l }{ \langle\eta \rangle^\alpha }\right)=O(Re_l),\\
 \langle\rho\rangle^\alpha\langle\bm u\rangle^\alpha\cdot\nabla\hat{\bm u}/\nabla\cdot(\langle \eta \rangle^\alpha (\nabla\hat{\bm u}+\nabla\hat{\bm u}^T ))=O\left(\frac{  \langle\rho \rangle^\alpha  \langle\bm u \rangle^\alpha l }{ \langle\eta \rangle^\alpha }\right)=O(Re_l).
\end{aligned}
\end{equation}
If the Reynolds number $Re_l$ is small enough, the terms on the left hand side of Eq.(\ref{eq:ap:simplifcaiton}) can be neglected. Then, Eq.(\ref{eq:ap:simplifcaiton}) can be reformulated as
\begin{equation}\label{eq:ap:simplifcaiton}
 -\nabla\hat{p}+\nabla\cdot(\langle \eta \rangle^\alpha (\nabla\hat{\bm u}+\nabla\hat{\bm u}^T ))
-\hat{\phi}\nabla \langle \mu \rangle^\alpha +\hat{\rho}\bm g
-\frac{\epsilon^{-1}}{ V}\int_{A_{\alpha\beta}}\left(-\hat{p}\bm I+\eta(\nabla\hat{\bm u}+\nabla\hat{\bm u}^T)  \right) \cdot\bm n_{\alpha\beta}dA  =0
\end{equation}

\section{Deviation equation for  the chemical potential}
In the work, we employed the assumption that the  chemical potential in each REV is constant, i.e, $\mu=\langle\mu \rangle^\alpha$. That is $\hat{\mu}=0$.
Subtracting Eq.(\ref{eq:pore-chemical}) from Eq.(\ref{eq:VA_Mu}) leads to
\begin{equation}\label{eq}
  \begin{aligned}
  \hat{\mu}=& -\nabla\cdot \kappa\nabla \hat{\phi}+\frac{1}{\epsilon}\nabla\cdot \frac{\kappa}{V}\int_{A_{\alpha\beta}}\hat{\phi}\bm ndA+\frac{1}{\epsilon}\frac{1}{V}\int_{A_{\alpha\beta}} \nabla\hat{\phi}\cdot \bm ndA \\
    & +12\lambda\hat{\phi}\langle\phi\rangle^\alpha \langle\phi\rangle^\alpha +12 \lambda \hat{\phi}\hat{\phi} \langle\phi\rangle^\alpha +4\lambda \hat{\phi}\hat{\phi}\hat{\phi}
    -12\lambda\langle\phi\rangle^\alpha\hat{\phi}-6\lambda\hat{\phi}\hat{\phi}+2\lambda\hat{\phi} \\
    & -12\lambda\langle \hat{\phi} \hat{\phi}\rangle^\alpha \langle \phi\rangle^\alpha-4\lambda\langle\hat{\phi}\hat{\phi}\hat{\phi}\rangle^\alpha+6\lambda\langle\hat{\phi}\hat{\phi}\rangle^\alpha.
  \end{aligned}
\end{equation}
Generally,  the spatial deviation order parameter is small compared to the volume averaged order parameter, i.e, $\hat{\phi}\ll \langle \phi\rangle^\alpha$~\cite{whitaker1998method}.
The terms involving $(\hat{\phi})^2$ can be neglected. Then, the simplified spatial deviation chemical potential reads
\begin{equation}\label{eq:ap:hatphi}
  \hat{\mu}=  -\kappa\nabla\cdot\nabla \hat{\phi}+\frac{\kappa}{\epsilon}\nabla\frac{1}{V}\int_A\hat{\phi}\bm ndA+\frac{\kappa}{\epsilon}\frac{1}{V}\int_A\nabla\hat{\phi}\cdot \bm ndA
   +12\lambda\hat{\phi}\langle\phi\rangle^\alpha \langle\phi\rangle^\alpha
    -12\lambda\langle\phi\rangle^\alpha\hat{\phi}+2\lambda\hat{\phi}.
\end{equation}
Inserting Eq.(\ref{eq:symplify_hat_phi}) to Eq.(\ref{eq:ap:hatphi}) leads to the constraint conditions for $\bm a_\phi$, $\bm b_\phi$ and $c_\phi$.

%
%

\section{Lattice Boltzmann method for the volume-averaged macroscopic equations}~\label{ap:lbm}
The lattice Boltzmann method  has emerged as a powerful too for  simulating complex fluid flow in porous media~\cite{guo2002lattice,guo2013lattice}.
For validation purposes, we develop a  multiphase lattice Boltzmann method for the REV-scale multiphase flow simulations.

The lattice Boltzmann equation with the single-relaxation-time can be expressed as
\begin{align}\label{eq:lbm}
h_i(\bm x+\bm e_i \delta t,t+\delta t)-h_i(\bm x,t)
&=-\frac{h_i(\bm x,t)-h_i^{eq}(\bm x,t)}{\tau_{h}}+H_i\delta t, \\
g_i(\bm x+\bm e_i \delta t,t+\delta t)-g_i(\bm x,t)
&=-\frac{g_i(\bm x,t)-g_i^{eq}(\bm x,t)}{\tau_{g}}+G_i\delta t,
\end{align}
where $h_i(\bm x,t)$ and $g_i(\bm x,t)$ are particle distribution functions for the
order parameter and the hydrodynamics fields, respectively. $\bm e_i$ is the discrete velocity in the $i$-th direction, $\delta t$ is the time step,
$\tau_h$ and $\tau_g$ are dimensionless relaxation times dependent on the
viscosity and mobility, respectively, and $H_i$ and $G_i$ are the discrete force term.
The equilibrium distribution functions $h_i^{eq}$ and $g_i^{eq}$ are respectively defined as
\begin{equation}\label{eq:CH_heq}
h_i^{eq}=\begin{cases}
 \epsilon \langle\phi\rangle^\alpha 
 +(\omega_i-1)\epsilon \zeta \langle \mu\rangle^\alpha,& \text{i=0}
 \\
 \omega_i\epsilon  \zeta\langle\mu\rangle^\alpha+ \epsilon \omega_i\frac{\bm e_i\cdot (\bm I +\langle c_\phi \bm B\rangle^\alpha)\cdot  \langle\phi\rangle^\alpha \langle \bm u\rangle^\alpha}{c_s^2},& \text{i$\neq$ 0}
\end{cases}
\end{equation}
\begin{equation}\label{eq:NS_geq}
\begin{aligned}
g_i^{eq}= & \omega_i \left[ \frac{\epsilon\langle p\rangle^\alpha }{c_s^2}
+
\epsilon \langle \rho\rangle^\alpha \left(\frac{\bm e_i\cdot \langle\bm u\rangle^\alpha}{c_s^2}
+\frac{\langle \bm u\rangle^\alpha \langle\bm u\rangle^\alpha :(\bm e_i\bm e_i-c_s^2\bm I)}{2c_s^4}\right)\right], 
\end{aligned}
\end{equation}
where $\omega_i$ is the weighting coefficient, $c_s$ is the sound speed, and $\zeta$ is a constant for a specified mobility to ensure that the relaxation time stays in a suitable range.
The source terms $H_i$ and $G_i$ can be expressed as
\begin{equation}\label{eq:CH_Force}
H_i=\omega_i\frac{\bm e_i\cdot \partial_t( \epsilon (\bm I+\langle c_\phi\bm B \rangle^\alpha)\cdot\langle \phi\rangle^\alpha \langle \bm u\rangle^\alpha )}{c_s^2}
+\omega_i \zeta \langle \mu\rangle^\alpha\bm e_i\cdot\nabla\epsilon
+\frac{\omega_i}{\tau_g-0.5}\frac{\bm e_i \cdot \epsilon \langle \bm B\cdot \langle \bm u\rangle^\alpha \bm b_\phi \rangle^\alpha \cdot \nabla\langle \phi\rangle^\alpha}{c_s^2}
\end{equation}
\begin{equation}\label{eq:NS_Force}
G_i=\omega_i\left[ \epsilon \langle \bm u\rangle^\alpha \cdot \nabla\langle\rho\rangle^\alpha
+ \frac{\bm e_i\cdot \bm F_1}{c_s^2}
+\frac{\langle \bm u\rangle^\alpha \bm F_2:(\bm e_i\bm e_i-c_s^2\bm I)}{c_s^4}
  \right].
\end{equation}
where $\bm F_1=\epsilon \langle \rho \rangle^\alpha \bm g-   \epsilon\langle \phi \rangle^\alpha \nabla \langle \mu \rangle^\alpha
-\epsilon^2\frac{\langle \eta \rangle^\alpha \langle \bm u\rangle^\alpha}{KK_r}-C_f\langle \rho \rangle^\alpha \epsilon^3 \frac{|\langle \bm u\rangle^\alpha|}{\sqrt{KK_r}}\langle \bm u\rangle^\alpha -\langle \eta\rangle^\alpha_{eff}(\nabla\langle\bm u\rangle^\alpha+(\nabla\langle\bm u\rangle^\alpha)^T )\cdot \nabla\epsilon
 +\langle p\rangle^\alpha\nabla\epsilon
 $ and $\bm F_2=\epsilon \langle \rho \rangle^\alpha \bm g-   \epsilon\langle \phi \rangle^\alpha \nabla \langle \mu \rangle^\alpha +c_s^2\epsilon\nabla\langle\rho \rangle^\alpha $.
 Finally,  the averaged quantities $\langle \phi\rangle^\alpha $,$\langle \bm u\rangle^\alpha $ and $\langle p\rangle^\alpha $ are calculated by
\begin{equation}\label{LBM_macroscopic}
\begin{aligned}
\epsilon \langle \phi\rangle^\alpha  &=\sum_i h_i, \\
\langle \bm u\rangle^\alpha &=\frac{2 \langle \bm u^*\rangle^\alpha }{C_1+ \sqrt{C_1^2+ 4C_2|\langle \bm u^*\rangle^\alpha|}}, \\
\epsilon \langle p\rangle^\alpha &=c_s^2\sum_i g_i+c_s^2\frac{\delta t}{2} \epsilon \langle\bm u\rangle^\alpha \cdot \nabla \langle \rho\rangle^\alpha.
\end{aligned}
\end{equation}
where $\langle \bm u^*\rangle^\alpha $ is a temporal velocity defined as
\begin{equation}\label{eq:temproal_velocity}
\langle \bm u^*\rangle^\alpha =\frac{1}{\langle\rho \rangle^\alpha}\left[\sum_i\bm e_i g_i+\frac{\delta t}{2}\left( \epsilon \langle \rho \rangle^\alpha \bm g -   \epsilon\langle \phi \rangle^\alpha \nabla \langle \mu \rangle^\alpha -\langle\eta\rangle^\alpha_{eff}(\nabla\langle\bm u\rangle^\alpha+(\nabla\langle\bm u\rangle^\alpha)^T)\cdot\nabla\epsilon
+\langle p\rangle^\alpha \nabla \epsilon \right)\right].
\end{equation}
\begin{equation}\label{eq}
C_1= \epsilon+ \frac{\delta_t}{2}\frac{\epsilon^2\langle \eta\rangle^\alpha}{\langle \rho \rangle^\alpha K_r K},
 \qquad  C_2=  \frac{\delta_t }{2}\frac{\epsilon^3 C_f  }{ \sqrt{K_r K}}
\end{equation}
Through the Chapman-Enskog analysis, the viscosity and the mobility are defined as
\begin{equation}\label{eq:vis_mob}
M_{\text{eff}}=c_s^2 \zeta(\tau_h-0.5)\delta t,\qquad  \langle\eta\rangle^\alpha_{\text{eff}}= \langle\eta \rangle^\alpha+ \langle \eta_d \rangle^\alpha =\langle \rho\rangle^\alpha  c_s^2(\tau_g-0.5)\delta t.
\end{equation}

In this study, we only consider two-dimensional cases, and the two-dimensional nine-velocity (D2Q9) LBE model is used, in which $\bm e_0=(0,0)$,
$\bm e_{i=1\ldots 4}=c{\cos[(i-1)\pi/2], \sin[(i-1)\pi/2] }$,
$\bm e_{i=5\ldots 8}=\sqrt{2}c{\cos[(2i-1)\pi/4], \sin[(2i-1)\pi/4] }$
and the corresponding weight coefficients are $\omega_0=4/9, \omega_{i=1\ldots 4}=1/9$,
and $\omega_{5\ldots8}=1/36$. The sound speed $c_s$ is defined as $c_s=c/\sqrt{3}$ where
$c=\delta x/\delta t$ with $\delta x$ being the lattice space.
In computations, the gradient operators are discretized with the second-order isotropic central scheme~\cite{zhang2019fractional}.
Taking the derivatives of the order parameter as example, the formulas can be written as
\begin{equation}\label{eq}
\begin{aligned}
\nabla \phi(\bm x)=\sum_{i\neq 0} \frac{\omega_i \bm e_i \phi(\bm x+\bm e_i\delta_t) }{c_s^2\delta_t},\\
\nabla^2 \phi(\bm x)=\sum_{i\neq 0} \frac{2\omega_i [\phi(\bm x+\bm e_i\delta_t)-\phi(\bm x)] }{c_s^2\delta_t^2}.
\end{aligned}
\end{equation}

\end{appendix}

\section*{ACKNOWLEDGEMENTS}
This work was supported by Fundamental Research Program of Shanxi Province (Grant No.202303021211160),  the National Natural Science Foundation of China(12472254).
and the Foundation of Stake Key Laboratory of Coal Combustion (No. FSKLCCA2401).
\section*{References}
\bibliography{mybib.bib}
\end{document}